\documentclass[aps,prev,twocolumn,preprintnumbers,floatfix,nofootinbib]{revtex4-1}
\newcommand{\D}{\mathrm{d}}
\newcommand{\be}{\begin{equation}}
\newcommand{\ee}{\end{equation}}
\newcommand{\bea}{\begin{eqnarray}}
\newcommand{\eea}{\end{eqnarray}}

\usepackage[utf8]{inputenc}
\usepackage{graphics}
\usepackage{mathrsfs}
\usepackage{amsmath, amsthm, amssymb}
\usepackage[dvipsnames]{xcolor}
\usepackage{epstopdf}
\usepackage[pdftex]{graphicx}
\usepackage{amssymb}
\usepackage{verbatim}
\usepackage{hyperref}
\usepackage{enumerate}
\usepackage{float}
\usepackage[caption = false]{subfig}
\usepackage[normalem]{ulem}  
\usepackage[capitalise]{cleveref}
\usepackage{csquotes}

\begin{document}

\title{Structure of Axion Miniclusters}
\author{David Ellis$^1$}
\email{david.ellis@uni-goettingen.de}
\author{David J. E. Marsh$^2$}
\email{david.j.marsh@kcl.ac.uk}
\author{Benedikt Eggemeier$^1$}
\email{benedikt.eggemeier@phys.uni-goettingen.de}
\author{Jens Niemeyer$^1$} 
\email{jens.niemeyer@phys.uni-goettingen.de}
\author{Javier Redondo$^{4,5}$}
\email{jredondo@unizar.es}
\author{Klaus Dolag$^{6,7}$}
\email{kdolag@mpa-garching.mpg.de}

\vspace{1cm}
\affiliation{$^1$Institut f\"{u}r Astrophysik, Georg-August Universit\"{a}t, Friedrich-Hund-Platz 1, D-37077 G\"{o}ttingen, Germany}
\affiliation{$^2$Theortetical Particle Physics and Cosmology, King's College London, Strand, London, WC2R 2LS, United Kingdom}
\affiliation{$^4$Departamento de F\'isica Te\'orica, Universidad de Zaragoza, 50009 Zaragoza, Spain}
\affiliation{$^{5}$Max-Planck-Institut f\"ur Physik (Werner-Heisenberg-Institut), F\"ohringer Ring 6, 80805 M\"unchen, Germany}

\affiliation{$^{6}$Universit\"{a}ts-Sternwarte, Fakult\"{a}t f\"{y}r Physik, Ludwig-Maximilians-Universit\"{a}t M\"{u}nchen, Scheinerstr.1, 81679 M\"{u}nchen, Germany}

\affiliation{$^{7}$Max-Planck-Institut f\"{u}r Astrophysik, Karl-Schwarzschild-Strasse 1, 85741 Garching, Germany}

\begin{abstract}
The Peak-Patch algorithm is used to identify the densest minicluster seeds in the initial axion density field simulated from string decay. The fate of these dense seeds is found by tracking the subsequent gravitational collapse in cosmological $N$-body simulations. We find that miniclusters at late times are well described by NFW profiles, although for around 80\% of simulated miniclusters a single power-law density profile of $r^{-2.9}$ is an equally good fit due to the unresolved scale radius.  Under the assumption that all miniclusters with an unresolved scale radius are described by a power-law plus axion star density profile, we identify a significant number of miniclusters that might be dense enough to give rise to gravitational microlensing if the axion mass is $0.2 \,\mathrm{meV}\lesssim m_a \lesssim 3\,\mathrm{meV}$. Higher resolution simulations resolving the inner structure and axion star formation are necessary to explore this possibility further.
\end{abstract}

\maketitle

\section{Introduction}

The axion is a theoretical elementary particle that can resolve the charge-parity (CP) problem~\cite{PhysRevLett.124.081803} of quantum chromodynamics (QCD) \cite{PecceiR.D.1977Ccit,PhysRevLett.40.279,PhysRevLett.40.223}, and provide a candidate to explain the observed~\cite{Aghanim:2018eyx} cosmological dark matter (DM)~\cite{DINE1983137,ABBOTT1983133,PRESKILL1983127}. Axion DM is of increasing interest both theoretically and experimentally~\cite{Chadha-Day:2021szb,Semertzidis:2021rxs}, with a number of constraints on its allowed properties~\cite{2016PhR...643....1M,Raffelt:2006cw,pdg}. 

The axion arises as the pseudo-Goldstone boson of a spontaneously broken global $U(1)$ symmetry, known as the Peccei-Quinn (PQ) symmetry. This symmetry is broken when the temperature of the universe falls below the vacuum expectation value, $v_{\rm PQ}$, of the PQ field, $\Phi=R e^{i\theta}$~\cite{PhysRevD.9.3357}, with $v_{\rm PQ}\propto f_a$, $f_a$ the ``axion decay constant'', and $\theta$ the axion field. Once the symmetry is broken, the axion field takes on a random ``misalignment angle'', $\theta_i(x)$, between zero and 2$\pi$.
The precise distribution of axion DM today depends on when in cosmological history the PQ symmetry is broken. There are two distinct scenarios.

In the present work, we focus on the case where PQ symmetry breaking occurs during the radiation dominated epoch (as opposed to the alternative, where it is broken during inflation). In this case, the present-day observable Universe is composed of many causally disconnected patches at the time of symmetry breaking. The random distribution of $\theta_i(x)$ leads to the formation of cosmic string topological defects\footnote{For simplicity, we consider only axion models with domain wall number equal to unity, such as the Kim-Shifman-Vainshtein-Zakharov (KSVZ) model~\cite{PhysRevLett.43.103,SHIFMAN1980493} and variations thereof~\cite{DiLuzio:2020wdo}.} due to the mapping between the vacuum $U(1)$ manifold, and space $\mathbb{R}^3$~\cite{Kibble_1976}.
When the axion mass becomes significant, $m_a(T)\gtrsim H(T)$ with $H$ the Hubble expansion rate, the $U(1)$ symmetry becomes strongly broken, and the strings decay as $\theta(x)$ relaxes to zero and oscillates everywhere.

%%%%%%%%%%%%%%%%%%%%%
\begin{figure}[t!]
\includegraphics[width=1\columnwidth]{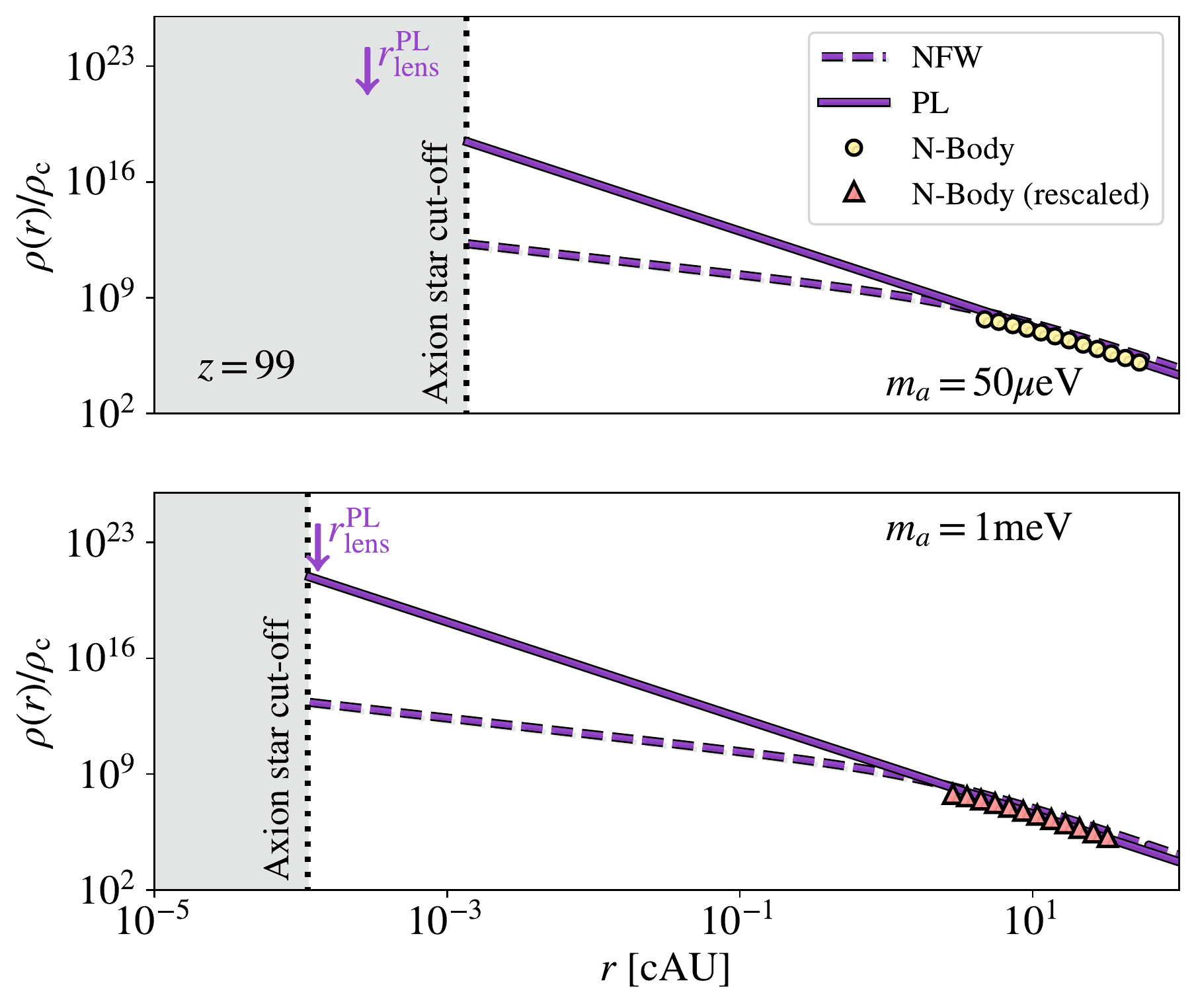}
\caption{\emph{Possible Minicluster Density Profiles}.
NFW profiles (dashed) are never dense enough to lens due to the large scale radius for the measured concentration relation (see \cref{fig:finalConc}). For power-law (PL) profiles there is a critical lensing radius, $r^{\rm PL}_{\rm len}$, which gives the minimum impact parameter of a minicluster along the line of sight in order to create above-threshold lensing amplification. This critical radius should be larger than the estimated axion star radius (indicated by the grey region). For $m_a=50\,\mu\text{eV}$ (upper panel) the estimated axion star radius is too large, while for $m_a=1\text{ meV}$, the estimated axion star radius can be significantly smaller. $N$-body simulations are rescaled to higher axion masses, and axion star radii are estimated from the measured velocity dispersion (see text for details).}
\label{fig:multi_ma_prof}
\end{figure}
%%%%%%%%%%%%%%%%%%%%%%%

The production of axions from strings~\cite{HARARI1987361} must be studied numerically, and the most recent simulations still lead to significant disagreements (caused by the numerical method, dynamic range, and extrapolations required) on the final abundance of axions and resulting axion mass required to explain the DM abundance~\cite{PhysRevD.83.123531,Klaer:2017ond,Gorghetto:2018myk,Buschmann:2019icd,Gorghetto:2020qws,Hindmarsh:2021vih,Buschmann:2021sdq,Hoof:2021jft}. Nonetheless, a prediction of this scenario is that string decay leaves large overdensities, $\delta$, in the axion field on the scale of the cosmological horizon size at string decay. The subsequent gravitational collapse of these overdensities leads to the formation of a class of DM substructure called ``axion miniclusters''~\cite{HOGAN1988228}.

The overdensities are of isocurvature type~\cite{Kolb:MCdens}, with a white noise spectrum on large scales and small-scale cut-off~\cite{Zurek:2006sy,khlopov_scalar}. They have a strongly non-Gaussian distribution, extending to large tails with $\delta\gg 1$~\cite{Kolb:1995bu,Vaquero:2018tib,Buschmann:2019icd,Ellis:2020gtq}. Large overdensities allow axion miniclusters to collapse under gravity in the radiation dominated epoch of cosmic history~\cite{Kolb:MCdens}, before hierarchical structure formation begins. Later, in the matter-dominated epoch, the miniclusters merge and form larger DM halos~\cite{Zurek:2006sy,Fairbairn:2017sil,Eggemeier:2019khm,Ellis:2020gtq}.

It is the purpose of the present work to determine the properties (mass, density profile, characteristic density etc.) of miniclusters and how they are related to the properties of the initial density field. We take initial density fields simulated using classical field theory~\cite{Vaquero:2018tib} analysed using the semi-analytic Peak-Patch algorithm~\cite{1996ApJS..103....1B, 1996ApJS..103...41B, 1996ApJS..103...63B,Stein:2018lrh,Ellis:2020gtq}, and compare to $N$-body simulations~\cite{Eggemeier:2019khm}. We apply our results to gravitational microlensing, but they can also be used to determine the minicluster distribution for direct and indirect detection~\cite{Kavanagh:2020gcy,Tinyakov:2015cgg,OHare:2017yze,Dokuchaev:2017psd,Edwards:2020afl}.

The density profile and mass of a minicluster is key to determining whether or not it is capable of producing an observable gravitational microlensing signature. After accounting for wave optics effects, only lenses with $M\gtrsim 10^{-11}M_\odot$ give rise to significant lensing magnification in the Subaru Hypersuprime Cam (HSC) microlensing survey~\cite{Niikura:2017zjd}. For an extended lens such as a minicluster, with density profile $\rho(r)$, shallow central densities prevent the formation of lensing caustics, and cut-off the magnification. Multiple different minicluster density profiles have been observed in simulation and proposed in theory. We demonstrate in Fig.~\ref{fig:multi_ma_prof} possible minicluster density profiles for two axion masses, indicating the important features to allow for microlensing. The density profiles are shown at $z=99$, the latest redshift available in simulation. The characteristic density is not expected to evolve at late times, though tidal stripping should be considered~\cite{Kavanagh:2020gcy}. We compare Navarro-Frenk-White (NFW) density profiles~\cite{10.1093/mnras/275.3.720, Navarro:1995iw, Navarro:1996gj} to single power law (PL) profiles $\rho(r)\propto r^{-2.9}$ (see Section~\ref{sec:density_profile_sim}). PL profiles give rise to lensing amplification $A>1.34$ (observationally dictated threshold) when they pass within a distance $r<r^{\rm PL}_{\rm len}$ along the line of sight (which we compute for lenses half way between Earth and M31).
%%%%%%%%%%%%%%%%%%%%%
\begin{figure*}[t!]
\includegraphics[width=1.5\columnwidth]{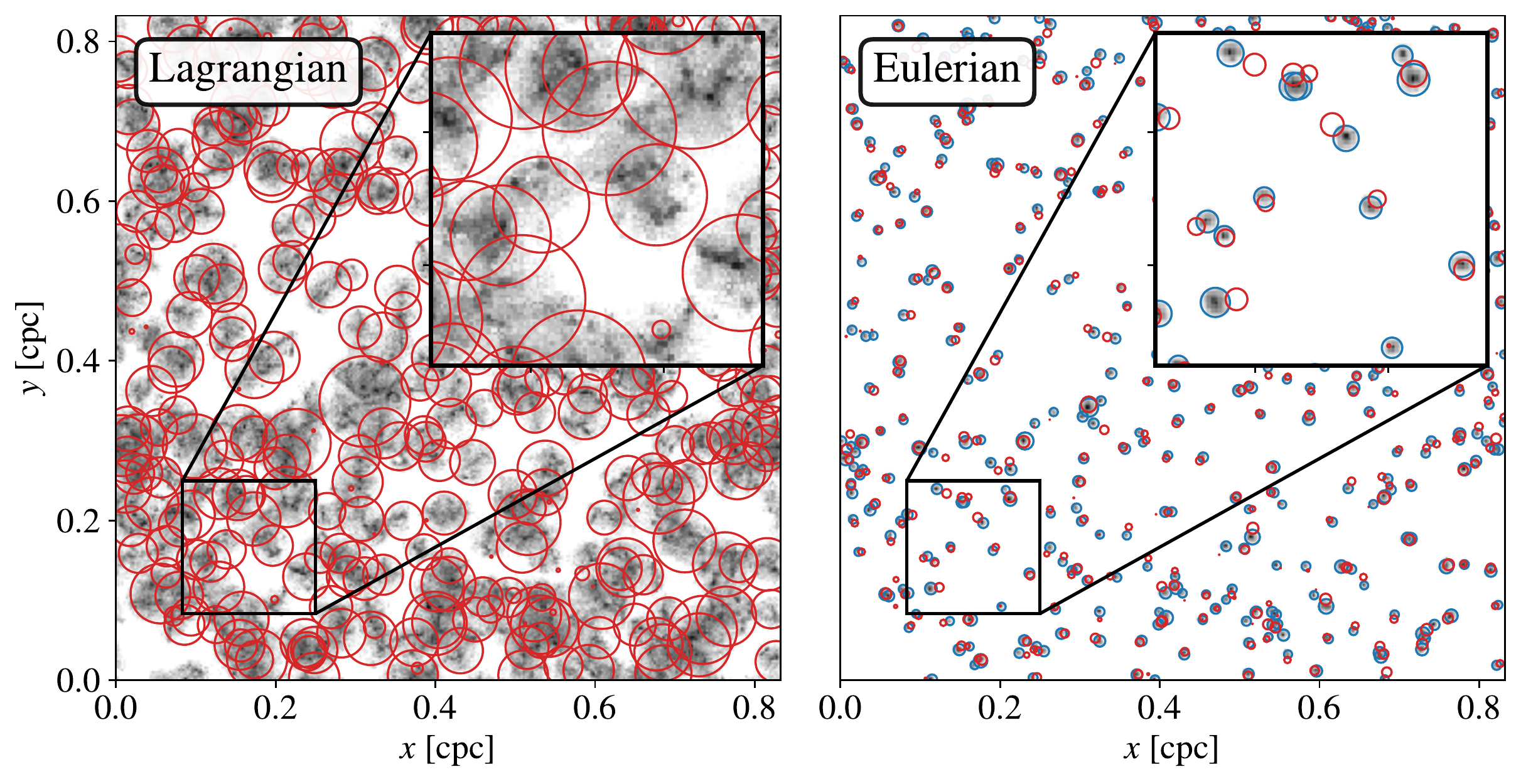}
\caption{\emph{Peak-Patch comparison with N-body}. Projection plot of the $N$-body particles (black) contained within the 400 largest halos located using the \textsc{Subfind} halo finder at a redshift $z = 3976.5$. Red circles indicate the most likely equivalent from the Peak-Patch data. The left panel shows Lagrangian coordinates, i.e. particles traced back to the initial density field, without displacement. The right panel shows the real-space (Eulerian) coordinates of the particles, which in Peak-Patch are found using first order Lagrangian displacements. In the right panel blue circles indicate the size and extent of the halos as found using \textsc{Subfind}. }
\label{fig:pp_nb_comp}
\end{figure*}
%%%%%%%%%%%%%%%%%%%%%

The key to microlensing is whether the outer steep power-law profile extends all the way to the central axion star (as observed in all simulations resolving axion star formation, e.g.~\cite{Eggemeier:2019jsu,Schive2014_Nature,Schive2014_PRL}), or whether the steep profile turns over with an inner scale radius before axion star formation (as observed in large scale $N$-body simulations that do not resolve axion star formation~\cite{Eggemeier:2019khm}). As we will show in the rest of this paper, there is a range of minicluster mass and axion mass where presently available simulations leave room for the possibility (within resolution limitations) of microlensing by miniclusters in the Subaru range. We discuss the necessary requirements of simulations that could settle this question. 

This paper is organised as follows. In Section~\ref{sec:bulkHalo} we describe our methods and previous results on the structure of minicluster halos, the mass function, and a comparison between $N$-body and semi-analytic methods. In \cref{sec:minicluster_seeds} we present new results that use the Peak-Patch method to tag and track dense ``minicluster seeds'' in $N$-body simulations, and characterise their properties. \cref{sec:lensing_results} describes the conditions for a minicluster to be a microlens candidate and tests a population of tagged minicluster seeds against them. We conclude in Section~\ref{sec:discussion}. The Appendices give details of our numerical methods, theory of minicluster formation, detailed results on the minicluster seed density profiles, assumed properties of axion stars, equations for microlensing, and the scaling relations used to translate our results to different axion masses. Throughout, we assume a $\Lambda$CDM standard cosmology with dark matter composed of the QCD axion alone. The Friedmann-Robertson-Walker scale factor is $a$, redshift is $z$, and we use units $\hbar=c=1$ for particle physics quantities. We consider two axion masses, $m_a=50\,\mu\text{eV}$ (simulated directly, approximately the minimum mass allowed in the relevant cosmology) and $m_a=1\,\mathrm{meV}$ (rescaled, where we find the most candidate microlenses).

\section{Minicluster Halos}
\label{sec:bulkHalo}

In Ref.~\cite{Ellis:2020gtq}, we studied the formation of miniclusters from the initial fields of Ref.~\cite{Vaquero:2018tib} using the Peak-Patch (PP) algorithm. In Ref.~\cite{Eggemeier:2019khm} we studied the formation of miniclusters from the same density field using $N$-body simulations. Here we compare the two studies in more detail and consider the properties of DM halos formed from hierarchical structure formation of miniclusters. 

The PP algorithm is an extended Press-Schechter (PS) model which solves the excursion set in real space \cite{1996ApJS..103....1B, 1996ApJS..103...41B, 1996ApJS..103...63B, Stein:2018lrh}. This semi-analytical method works on the linearly evolved initial density field. The density field is smoothed on a hierarchy of scales, a cell is then considered to be collapsed if its overdensity is above some redshift dependent threshold when smoothed on any of these scales. The masses of halos are predicted by performing a radial integration centred on local peaks in this overdensity until the total average overdensity is equal to the threshold. The final positions of each halo are calculated using first-order Lagrangian displacements (see Appendix~\ref{appendix:pp} for more details).

We modified PP in Ref.~\cite{Ellis:2020gtq} to work on the minicluster initial density field and account for the effects of a radiation dominated background. A comparison between this modified PP and an $N$-body simulation is presented in Fig.~\ref{fig:pp_nb_comp}. The left and right panels show projection plots of the $N$-body particle density in Lagrangian and Eulerian coordinates (i.e. coordinates on the initial density field, and the real-space final coordinates~\cite{1970A&A.....5...84Z,Bouchet:1994xp}), respectively, for the 400 largest halos found in our simulations using the \textsc{Subfind} halo finder~\cite{Subfind, Dolag2009}. The Eulerian locations of these halos as found by \textsc{Subfind} are shown in blue in the right panel. The most likely equivalent halos in the PP data are shown in red in both sets of coordianates. We see that the PP estimates are in close agreement with $N$-body, particularly in Lagrangian coordinates. This makes sense since PP works on the initial density field. In Eulerian space, the sizes and masses of the halos are well estimated by PP, while the locations are systematically offset. This can be understood since we use only first-order Lagrangian displacements to move from Lagrangian space to the predicted Eulerian coordinates. These results confirm the accuracy of our modifications to PP described in Ref.~\cite{Ellis:2020gtq} for producing accurate halo catalogues in minicluster cosmologies with isocurvature initial conditions, large non-Gaussianities, and collapse during the radiation epoch (a similar comparison for standard $\Lambda$CDM cosmologies was shown in Ref.~\cite{Stein:2018lrh}). 
    
\begin{figure}[t!]
    \centering
    \includegraphics[width=1\columnwidth]{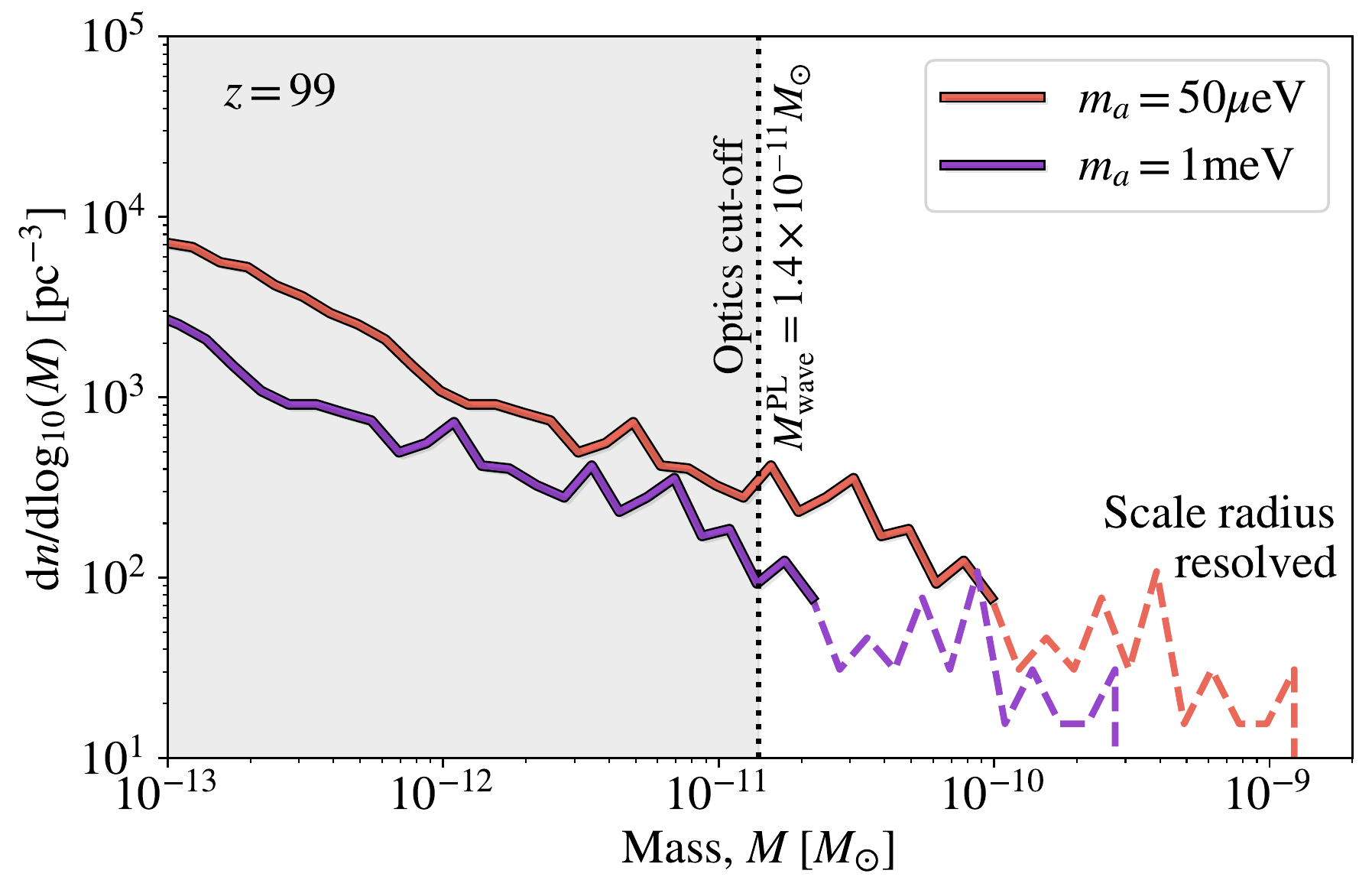}
    \caption{\emph{Minicluster Mass Function at $z=99$}: Dashed lines indicate the portion of the mass function with resolved scale radius, i.e. known NFW halos. The vertical line indicates the minimum mass that a pure power-law halo must have in order to produce significant lensing amplification after accounting for wave optics effects and assuming a power-law profile (see Eq.~\ref{eqn:wave_optics_PL_halo_mass}).}
    \label{fig:HMF}
\end{figure}

In Ref.~\cite{Ellis:2020gtq}, we used PP to calculate the minicluster Halo Mass Functions (HMFs) at a range of redshifts. These were found to be in good agreement with $N$-body simulations performed on the same initial density field in Ref.~\cite{Eggemeier:2019khm}. As discussed in Ref.~\cite{Ellis:2020gtq}, PP and $N$-body HMFs disagree for $z\lesssim 600$ due to box-scale non-linearities. This affects the exact minicluster number statistics at very small and very large halo mass.
Nonetheless, the inner structure of individual halos is unaffected and we can study the density profiles of (sub)halos
with masses down to $\sim 10^{-13}\,M_{\odot}$ at low redshifts.\footnote{This lower limit was chosen to guarantee that (sub)halos consist of a sufficient number of particles to study their density profiles.} Fig.~\ref{fig:HMF} shows the $N$-body HMF at $z=99$ for the simulated $m_a=50\,\mu\text{eV}$, and also the rescaled HMF to $m_a=1\,\mathrm{meV}$.

\begin{figure}[t!]
\includegraphics[width=\columnwidth]{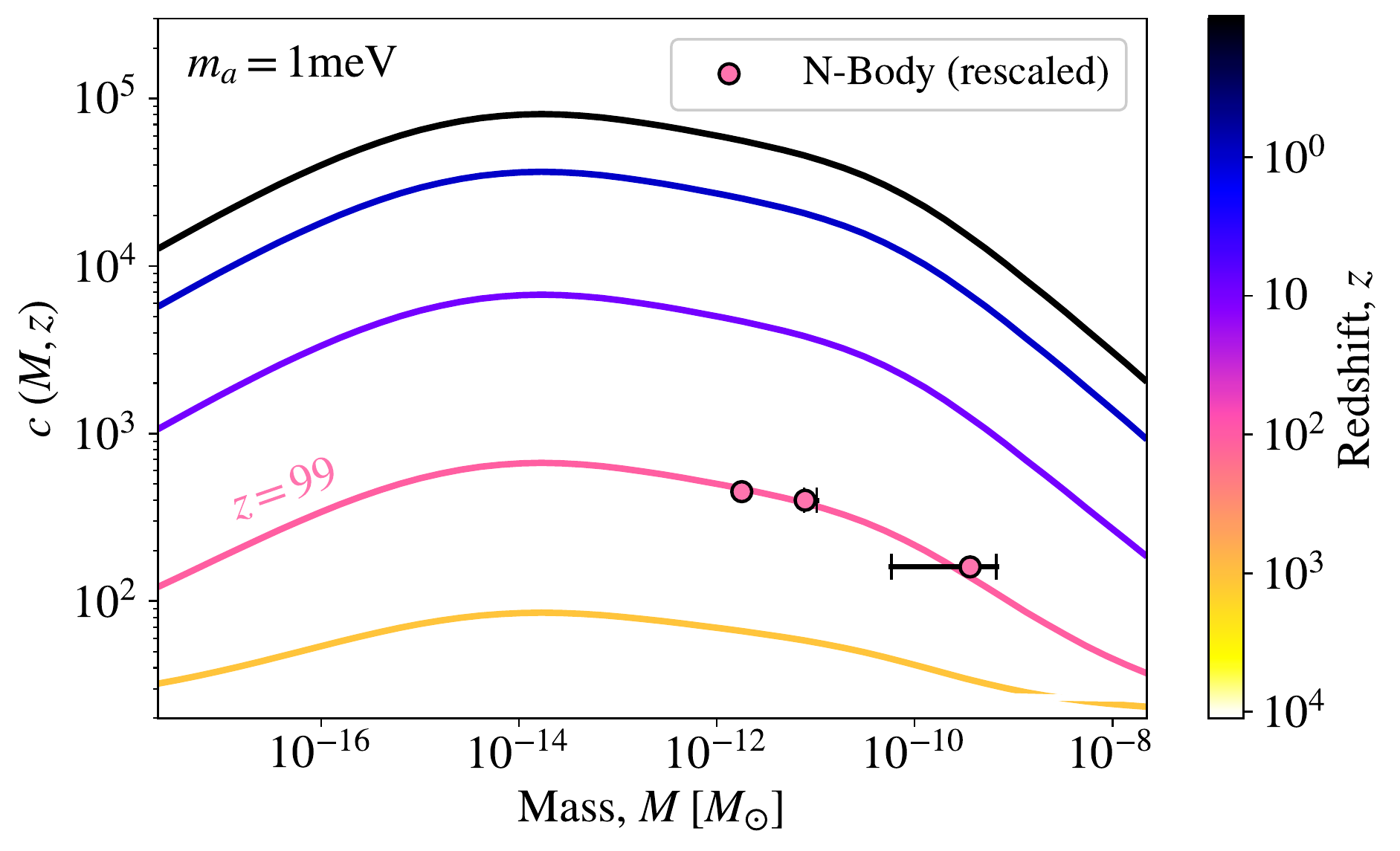}
\caption{\emph{Minicluster Concentrations}. Press-Schechter predicted minicluster concentrations at different redshifts, assuming Gaussianity and NFW profiles. Points show $N$-Body results used for normalisation. All masses were rescaled to the axion mass $m_a = 1\text{ meV}$.}
\label{fig:finalConc}
\end{figure}

In Ref.~\cite{Ellis:2020gtq} we also used PP to study minicluster concentrations. Our $N$-body simulations in Ref.~\cite{Eggemeier:2019khm} have shown that minicluster halos have NFW profiles given by
\begin{equation}
    \rho(r) = \frac{\rho_0}{r/r_s\left(1+r/r_s\right)^2}\,,
    \label{eq:nfw_profile}
\end{equation}
where $\rho_0$ and $r_s$ are the scale density and radius respectively. The concentration is defined as the ratio of the virial and scale radius~\cite{Navarro:1996gj}
\begin{equation}
    c = \frac{R_{\mathrm{vir}}}{r_s}\,.
\end{equation}
The NFW model sets the halo's scale density proportional to the density of the universe when the halo collapsed~\cite{Navarro:1996gj}. The NFW model defines the collapse redshift, sometimes also called the formation redshift, to be the redshift at which for some halo half of its final mass $M_{\mathrm{final}}$ is contained within progenitors of a mass larger than $fM_{\mathrm{final}}$ where $f$ is some fraction. Using PP, we built merger trees for each halo to calculate their collapse redshift $z_{\mathrm{col}}$. Their scale density at a later redshift $z$ can then be predicted via 
\begin{equation}
    \delta_s(z, \kappa) = \kappa(f) \frac{\bar\rho(z_\mathrm{col})}{\bar\rho(z)}\,,
    \label{eq:NFWConc1}
\end{equation}
where $\bar\rho$ is the average density of the universe and $\kappa(f)$ is a constant of proportionality obtained from a fit to the $N$-body simulations. The concentration parameter can then be calculated by solving
\begin{equation}
    \delta_s = \frac{200}{3}\frac{c^3}{\ln(1+c) - c/(1+c)}\,.
    \label{eq:NFWConc2}
\end{equation}
We calibrate the minicluster $c(M)$ from PP and Press-Schecter to the $N$-body results at $z=99$, which fixes $\kappa(f)$. For large concentrations, so long as the collapse redshift does not change significantly, the concentration evolves proportionally to the scale factor, $a=1/(1+z)$. We are therefore able to project our results to $z=0$. The resulting $c(M)$ is shown in Fig.~\ref{fig:finalConc}.

%%%%%%%%%%%%%%%%%%%%%%%%%%%%%%
%%%%%%%%%%%%%%%%%%%%%%%%%%%%

\section{Tracking Minicluster Seeds in $N$-body Using PP}\label{sec:minicluster_seeds}

In the last section, we characterised minicluster halos, formed via hierarchical mergers and described by an NFW density profile.  However, these do not represent the densest ``minicluster seeds'' which may survive from the early, $z>z_{\rm eq}$, epoch of structure formation characterised by peak collapse and infall~\cite{Eggemeier:2019khm,Ellis:2020gtq}. We now demonstrate how to identify collapsed peaks at $z_{\rm eq}$ using PP, and then follow the fate of these selected collections of particles at late times in $N$-body simulation, in order to determine the survival rate of dense seeds, and their density profiles. 

\subsection{Initial Overdensity, Mass, and Survival}
\label{sec:init_overdens_mass_survival}
In the theory of the spherical collapse of miniclusters, their final density is determined by the ``initial overdensity'' (see Appendix~\ref{sec:infall}). The initial overdensity depends on a scheme for filtering the density field. We use physically motivated filtering based on the redshift of collapse.

\begin{figure}[t!]
\includegraphics[width=\columnwidth]{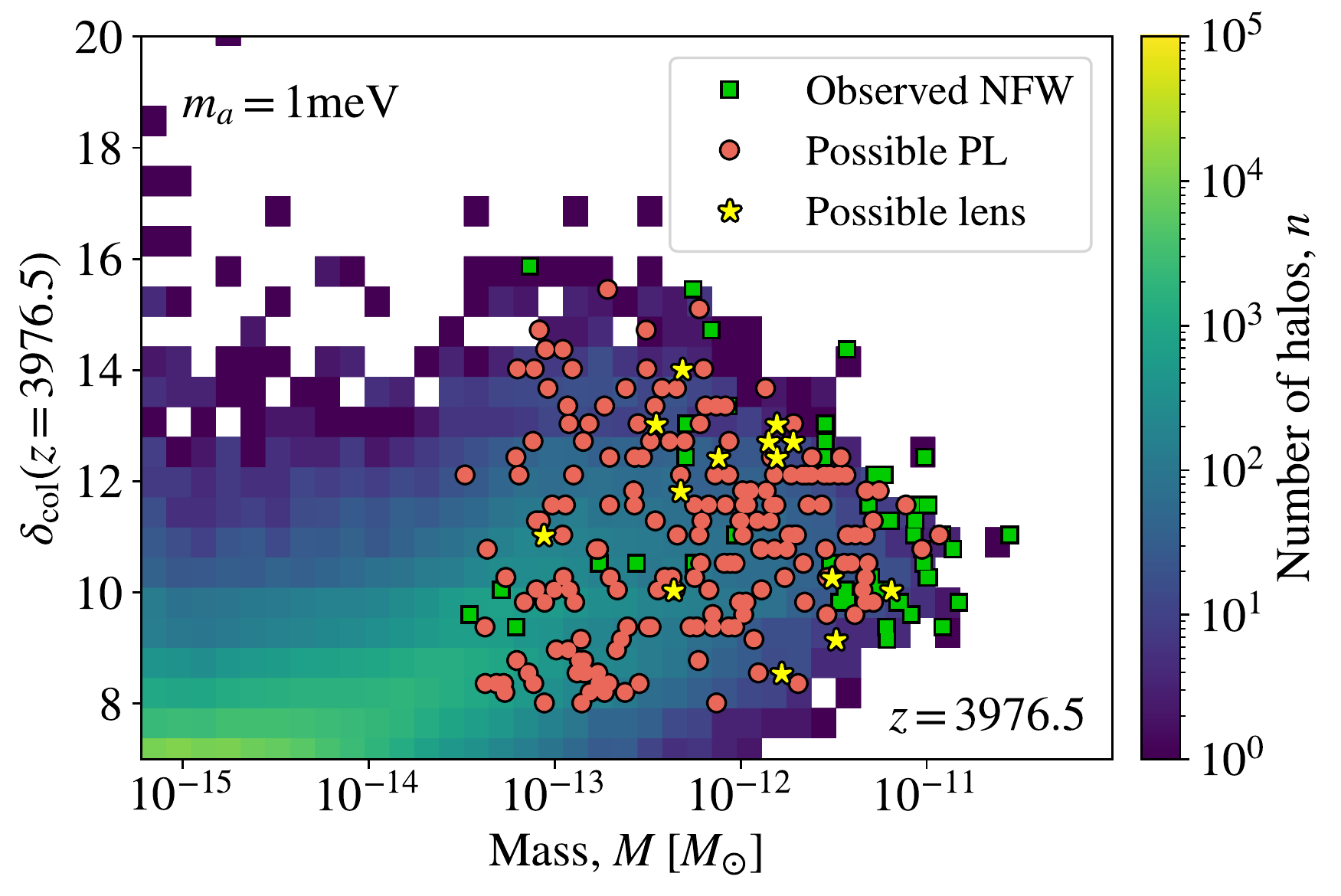}
\caption{\emph{Minicluster distribution}. We use Peak-Patch to identify all miniclusters at $z=3976.5$ and measure their mass and overdensity at collapse redshift, $\delta_{\rm col}$ (see \cref{eq:init_del}). Overlaid points show the sample of miniclusters tracked in $N$-Body simulations, which survived until $z=99$ (index matching fraction $f_\mathrm{m}>0.5$). All masses were rescaled to the axion mass $m_a = 1\,\mathrm{meV}$.}
\label{fig:delta_vs_mass}
\end{figure}

Using PP, we build merger trees as done in Ref.~\cite{Ellis:2020gtq}, and using the collapse redshift we assign to each halo at $z=3976.5$ an initial overdensity $\delta_{\mathrm{col}}(z)$ via \cref{eq:init_del}. Samples of minicluster seeds are randomly drawn from the distribution shown in \cref{fig:delta_vs_mass} such that the samples have a uniform distribution in $\delta_{\rm{col}}$ with $\delta_{\rm{col}}>8$ and a uniform distribution in $\log_{10} M$ with $M > 5\times10^{-14} M_{\odot}$ (chosen as an arbitrary low mass cut). PP defines these miniclusters within the Lagrangian sphere on the initial  ($z\sim10^6$)  density field. This procedure allows us to locate these particles again in later snapshots, and to study the properties and evolution of minicluster seeds. We then track the particle IDs in the $N$-body simulations of Ref.~\cite{Eggemeier:2019khm}, specifically a total number of 361 miniclusters to $z=3976.5$ and 1029 miniclusters to $z=99$ (limited by computational time in post-processing). We then note the IDs of particles in the initialparticle distribution contained  of these objects. 

We now want to determine what has happened to these collections of particles in the $N$-body snapshots. We begin by determining the point of maximum density in each minicluster seed using a kernel density estimation (KDE) algorithm. We then ``collect'' these particles by working radially outwards from the centre until the predicted final mass is found. We estimate this mass by assuming that from $z=3976.5$ the halos only accrete matter from the background. As such, we predict that the mass will evolve as~\cite{Berezinsky:2014wya}:
\begin{equation}
    M = M(z'=3976.5)\left(\frac{1+3976.5}{1+z}\right)\,.
    \label{eq:massGrowth}
\end{equation}

Not all of the objects we tag at $z_{\rm eq}$ will survive until $z=99$. Many will be completely disrupted through mergers with larger halos. We define a survival metric as the ``index matching fraction'' $f_\mathrm{m}$. This is the fraction of particles that were marked in the initial density field which are in a sphere around the maximum density containing the expected mass at a later redshift. If $f_\mathrm{m}\approx 1$, then most of the original mass remains close to the density peak at a later redshift, while if $f_\mathrm{m}\ll 1$, then the originally tagged particles have dispersed far from one another.

\begin{figure}
\includegraphics[width=\columnwidth]{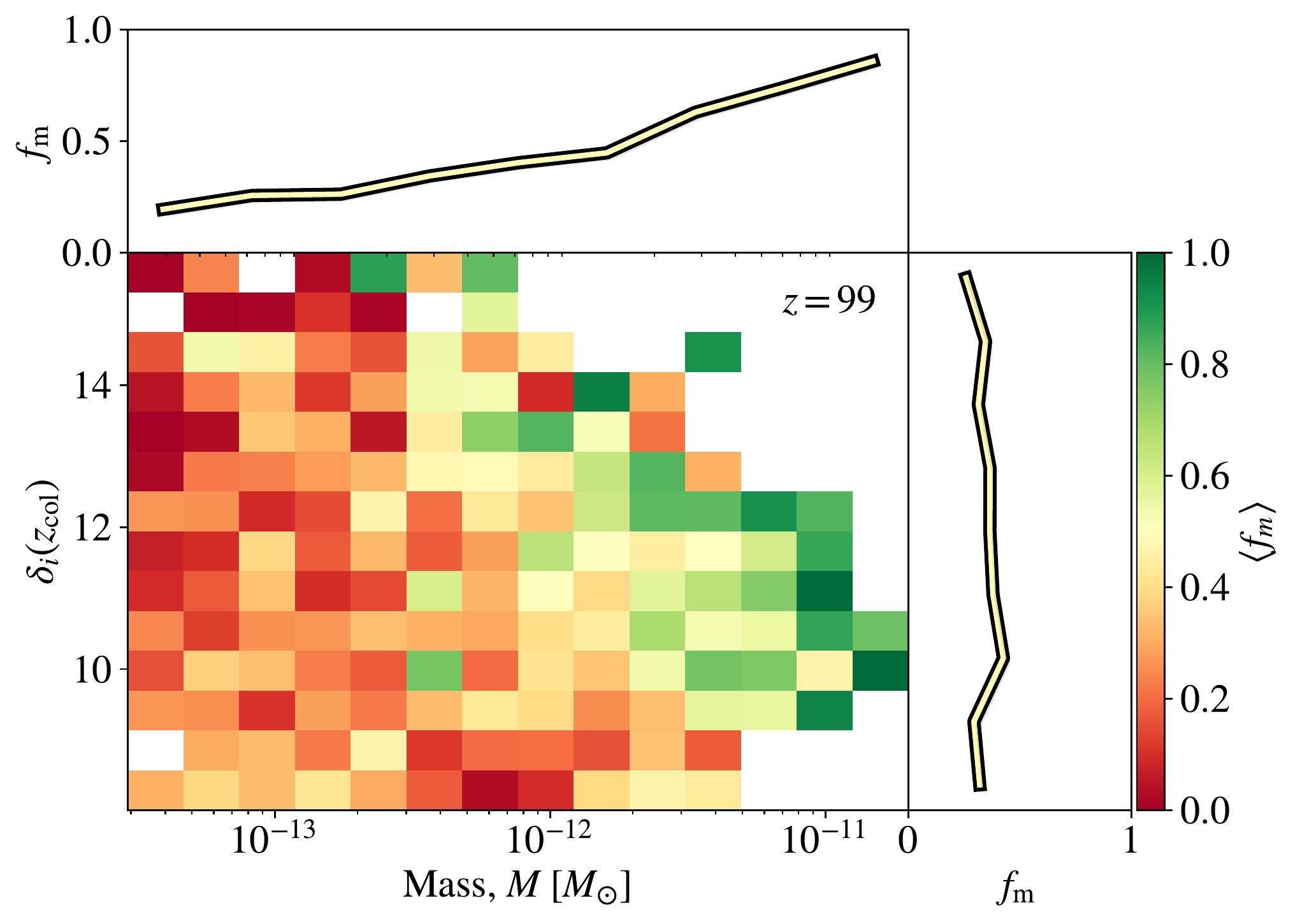}
\caption{\emph{Trends in minicluster survival}. Two-dimensional histogram for index matching fraction $f_\mathrm{m}$ as a function of mass and $\delta_i(z_\mathrm{col})$ for all sample halos at $z=99$. The top panel shows $f_\mathrm{m}$ as a function of sample mass $M$ and $f_\mathrm{m}$ as a function of $\delta_i(z_\mathrm{col})$ is shown on the right. This figure uses the simulated value $m_a=50\,\mu\text{eV}$.}
\label{fig:survival}
\end{figure}

The index matching fraction of our halo sample is shown in \cref{fig:survival} as a function of halo mass and initial overdensity. We use this quantity to distinguish the density profiles of our sample miniclusters. We observe a strong correlation between $f_\mathrm{m}$ and mass.
This is as we would expect, since the most massive objects have the deepest potential wells, and have undergone fewer major mergers. We also observe higher $f_\mathrm{m}$ at higher $\delta_i$ for masses $\gtrsim 3\times 10^{-13}\,M_\odot$, consistent with the hypothesis that a larger $\delta_i$ leads to a minicluster of higher average density after virialisation, and that higher densities are more resistant to tidal stripping~\cite{Kolb:MCdens,Dokuchaev:2017psd,Tinyakov:2015cgg}. 
%\tkBE{Added mass scale since overall correlation between $\delta_i$ and $f_m$ does not seem to exist. Right panel of \cref{fig:survival} shows even a reduced $f_m$ at largest $\delta_i$.}
 
\begin{figure}[t]
    \includegraphics[width=\columnwidth]{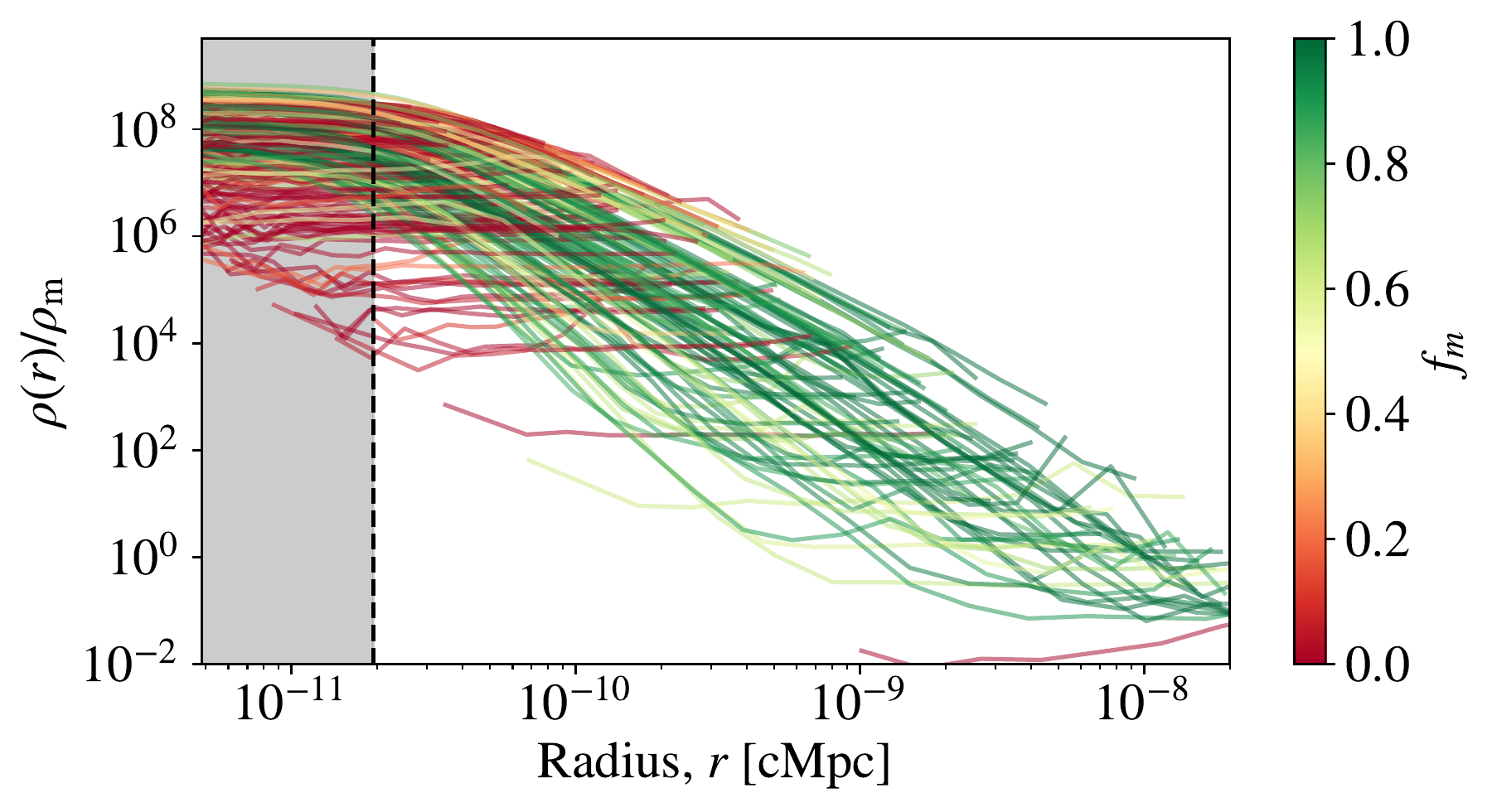}
    \caption{\emph{Demonstration of survival metric.} Random selection of 300 halos from the $(M,\delta_i)$ plane followed to $z=99$ in $N$-body simulations. We pot the radial density profile evaluated in spheres from the point of maximum density of the tagged particles. Those miniclusters with a high index matching fraction $f_\mathrm{m}$ have much more well-defined density profiles, indicating that $f_\mathrm{m}$ is a good metric for the minicluster surviving the merger process. The grey shaded region indicates the softening cut-off. This figure uses the simulated value $m_a=50\,\mu\text{eV}$.}
    \label{fig:match_frac}
\end{figure}

We plot density profiles centred on the maximum density of tagged particles for all objects in our sample in Fig.~\ref{fig:match_frac}, colour coded by the value of $f_\mathrm{m}$. We observe that halos with a large $f_\mathrm{m}$ tend to have a much more well-defined density profile. 
This demonstrates that a large value of $f_\mathrm{m}$ corresponds to an object that remains gravitationally bound.

%%%%%%%%%%%%%%%%%%%%%%%%
%%%%%%%%%%%%%%%%%%%%%

%%%%%%%%%%%%%%%
%%%%%%%%%%%%%%%%%%

%%%%%%%%%%%%%%%%%
%%%%%%%%%%%%%%%%%%

\subsection{Density Profiles}\label{sec:density_profile_sim}

We now analyse the radial density profile of our selected miniclusters, comparing them both to NFW and single power-law models. The analysis of the radial density profiles is restricted to the spatial resolution of the $N$-body simulations. The lower limit is expressed in terms of the numerical softening length which was set to $1\,\mathrm{AU}/h$ in comoving units in the simulations from Ref.~\cite{Eggemeier:2019khm}. Following the resolution studies in Refs.~\cite{Power:2002sw,Zhang:2018nqh}, we only consider scales larger than $4\,\mathrm{AU}/h$ when analysing the radial density profiles of miniclusters in order to be safely above the softening length.

Note that by evolving our minicluster sample from $z=3976.5$ to $z=99$ assuming only accretion from the background according to \cref{eq:massGrowth} we under-predict some of their actual virial masses at $z=99$ obtained from the N-body simulations. This implies that these objects gain a significant proportion of their mass from mergers as opposed to only from accretion. Therefore, we extrapolate their density profiles as a power-law to larger $r$ to match their actual virial radii such that the enclosed halo overdensity is $\Delta=200$.

\begin{figure*}
    \centering
    \includegraphics[width=\textwidth]{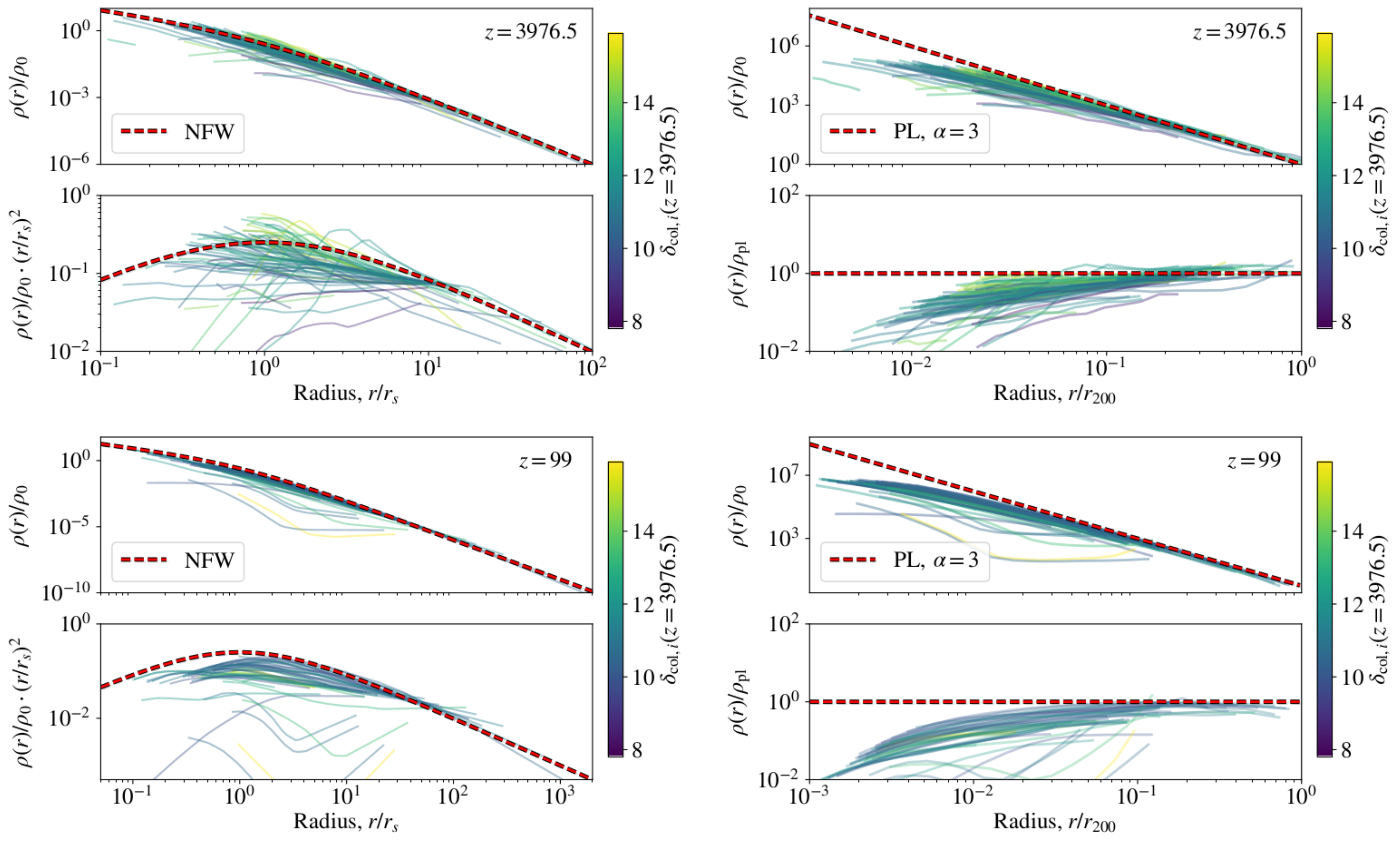}
    \caption{\emph{Minicluster density profiles with resolved scale radius.} In total,  76 samples with a matching-fraction $f_m>0.5$ are shown at $z=3976.5$ (top) and 48 samples with a matching-fraction $f_m>0.75$ at $z=99$ (bottom). The line colours are given by $\delta_{\rm{col, i}}(z = 3976.5)$. \emph{Left}: NFW prediction calibrated from $c(M)$, where the shown density profiles are additionally multiplied with $r/r_s$ to highlight the turnover of the NFW profile at $r_s$. \emph{Right}: Power-law density profile with a slope parameter of $\alpha=3$ and its deviation from the sample. This figure uses the simulated value $m_a = 50\,\mu\mathrm{eV}$.}
    \label{fig:densprofile_resolved}
\end{figure*}

\begin{figure*}
    \centering
    \includegraphics[width=\textwidth]{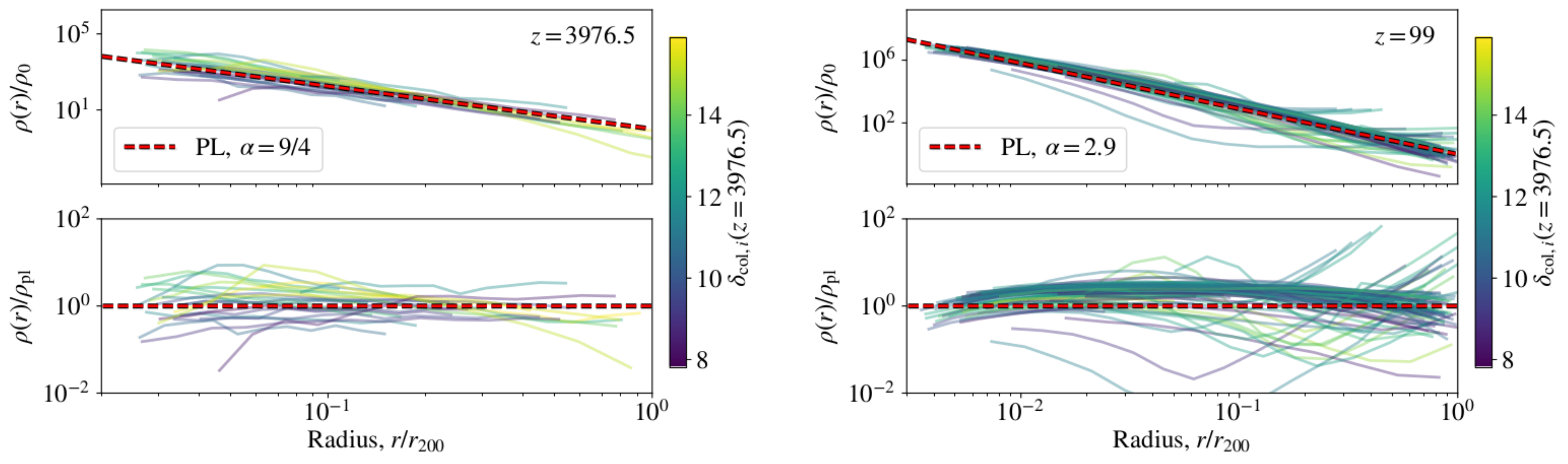}
    \caption{\emph{Minicluster density with unresolved scale radius.} In total, 25 samples with a matching-fraction $f_m>0.5$ are shown at $z=3976.5$ (left) and 100 samples with a matching-fraction $f_m>0.75$ at $z=99$. The density profiles are compared to single power-law fits with slope-parameters of $\alpha=9/4$ and $\alpha=2.9$, respectively. This figure uses the simulated value $m_a = 50\,\mu\mathrm{eV}$.}
    \label{fig:densprofile_unresolved}
\end{figure*}

We compare the density profiles to NFW and single power-law models at $z=99$ and $z=z_{\rm eq}$ in \cref{fig:densprofile_resolved,fig:densprofile_unresolved}. For NFW profiles we plot $\rho(r)/\rho_0$ against $r/r_s$, and for PL profiles we plot $\rho(r)/\rho_0$ against $r/R_{\rm{vir}}$. We can therefore compare the profiles of our whole sample to a single ``predicted'' curve. Since only around $20\%$ of our sample at $z=99$ have a predicted scale radius that is spatially resolved, we study the density profiles of miniclusters with resolved and unresolved scale radius separately from each other in \cref{fig:densprofile_resolved} and \cref{fig:densprofile_unresolved}, respectively. In both cases, there is no visible correlation between the initial overdensity of the miniclusters and their density profiles.

We find that the density profiles in \cref{fig:densprofile_resolved} are well described by an NFW profile at $z=99$ with a visible turnover at the scale radius. The turnover is less pronounced in the density profiles at $z=3976.5$ and we thus compare them also with a single power-law density profile with a slope parameter of $\alpha=3$. As expected from the general agreement with the NFW profile, the power-law profile well describes the density profiles for large $r$. However, at smaller radii there are increasing deviations confirming that the density profiles become shallower favouring an NFW profile. Increasing deviations between the single power-law profile and the density profiles at small radii can also be observed at $z=99$, confirming that they can be well described by NFW profiles. 

Since the predicted scale radii of the remaining $80\%$ of our sample at $z=99$ are not spatially resolved, it is possible that these halos instead have a single power-law extending to smaller radii. For the halos with unresolved $r_s$, we performed a least-squares fit for a single power-law  in the range between the resolution cut-off and the virial radius, $4\,\mathrm{AU}/h < r < R_\mathrm{vir}$, and found that the average slope is $\alpha \approx 2.9$. This power-law is in good agreement with the density profiles at $z=99$, as shown on the right-hand side of \cref{fig:densprofile_unresolved}. At matter-radiation equality, the density profiles of the miniclusters with unresolved scale radius are in agreement with an average slope of $9/4$ which means that they cannot be described by an NFW profile, even at large radii. The slope of $9/4$ is in agreement with the theory of self-similar infall (see Appendix~\ref{sec:infall}).

\subsection{Estimating Axion Star Properties}
\label{sec:stars_estimates}

\begin{figure}[t!]
\includegraphics[width=1\columnwidth]{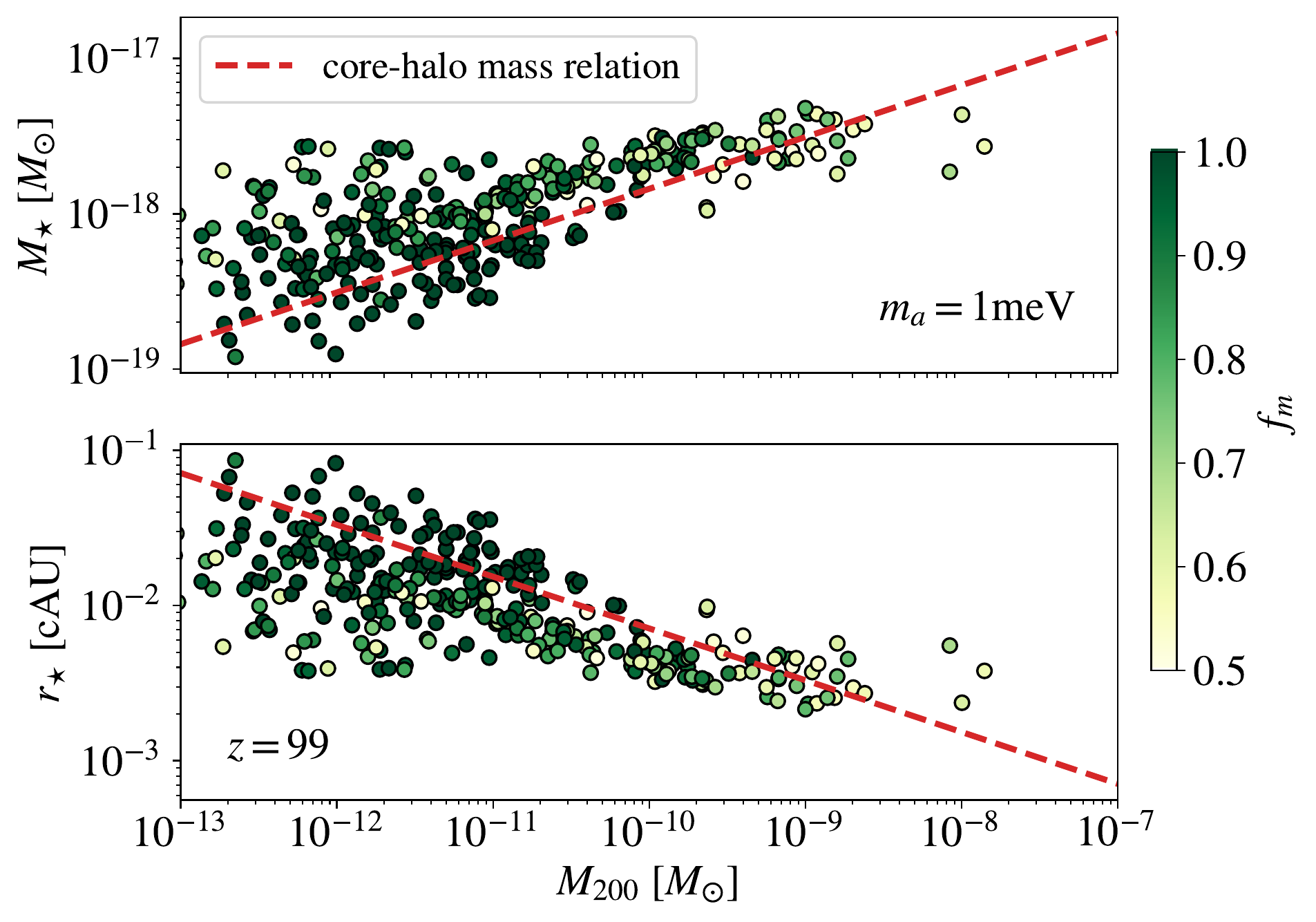}
\caption{\emph{Axion star mass and radius.} Estimated by calculating $v_{\rm{vir}}$ for $N$-body halo samples with $f_\mathrm{m}>0.5$ and substituting into \cref{eq:axionstar_mass,eq:axionstar_halfmassradius}, respectively. The $N$-body data were rescaled to the axion mass $m_a=1\,\mathrm{meV}$.}
\label{fig:massesAndradii}
\end{figure}

Axion stars are expected to form at the centres of miniclusters~\cite{Eggemeier:2019jsu}. Our $N$-body simulations~\cite{Eggemeier:2019khm}, on the other hand, are not able to resolve wave-like dynamics and axion star formation. 
However, we can estimate the axion star masses and radii using known scaling relations (see Appendix~\ref{appendix:star} for details). Ref.~\cite{Eggemeier:2021smj} showed that the star mass distribution estimated in this way from particle data accurately reproduces the results of simulations resolving the wavelike dynamics.

Using \cref{eq:axionstar_mass_velocity}, we can relate the virial velocity of a halo to the mass of the axion star. For typical virial velocities of $v_\mathrm{vir}\sim 0.1\,\mathrm{m}/\mathrm{s}$ it can be expressed as
\be
    M_{\star} = 2.08\times10^{-16} \left( \frac{1\,\mathrm{meV}}{m_a}\right)\left( \frac{v_{\mathrm{vir}}}{0.1\,\mathrm{m}\,\mathrm{s}^{-1}} \right) M_\odot\,.
    \label{eq:axionstar_mass}
\ee
Using the soliton density profile given in Eq.~\eqref{eqn:soliton_profile}~\cite{Schive2014_Nature,Schive2014_PRL}, its half-mass radius is given by 
\be
    r_\star = 4.95\times10^{-12} \left( \frac{1\,\mathrm{meV}}{m_a}\right)^2 \left( \frac{10^{-16}M_{\odot}}{M_{\star}}\right)\,\mathrm{pc}\,.
    \label{eq:axionstar_halfmassradius}
\ee
Calculating this for our particle samples for minicluster seeds gives the distributions shown in Fig.~\ref{fig:massesAndradii}. 

We observe significant scatter in the distribution. The scatter in the distribution allows some halos to have heavier, and thus more compact, axion stars, than a simple power-law relation between halo mass and axion star mass would predict. This is of key importance for microlensing, as we now show. We discuss the possible relevance of this scatter for the ``core-halo mass relation''~\cite{Schive2014_PRL,Chan:2021bja} in Section~\ref{sec:discussion}.

\section{Microlensing by Miniclusters}\label{sec:lensing_results}

\subsection{Conditions for Minicluster Microlensing}
\label{sec:lensing_conditions}

Gravitational microlensing is briefly reviewed in Appendix~\ref{appendix:micro}.  Microlensing occurs when a lens passes within a ``tube'' between the source star and the observer, defined by the minimum impact parameter required to produce an observed amplification of $A=1.34$~\cite{Griest:1990vu}. For a point mass, the tube radius is the Einstein radius, $R_E$, while for an extended lens the tube radius is rescaled by a factor of $\mathcal{R}$ (see e.g. Ref.~\cite{Fairbairn:2017sil}), so that the lensing tube has radius $\xi_c=\mathcal{R} R_E$. For a lens with density profile $\rho(r)$, in order for $A>1.34$, one requires the slope of the density profile to increase with a power-law steeper than $r^{-1}$ within the lensing tube (see Eqs.~\ref{eq:extLens},\ref{eq:extLens2}).  For NFW profiles, this implies that lensing can only occur if $r_s\lesssim R_E$~\cite{Fairbairn:2017sil}.

Axion stars have shallow density profile slopes within the half-mass radius $r_\star$. The axion sar acts as a small-scale cut-off even to a PL profile. Thus, for the same reason that NFW profiles lens only if $r_s\lesssim R_E$, a PL profile can only lens if $r_\star<\xi_c$, where $\xi_c$ is the rescaled lensing tube radius for the PL profile.

Furthermore, for an optical microlensing survey with a typical filter wavelength of $\lambda = 6210\,\text{\AA}$, the mass contained within the lensing tube, $M_{\rm eff}$, must exceed a critical mass, $M_{\rm min}\approx 3 \times 10^{-12}\,M_{\odot}$ such that wave optics effects do not suppress the magnification (see e.g. Ref.~\cite{Niikura:2017zjd}). 

We show in \cref{appendix:micro} that miniclusters described by an NFW profile for all $M$ cannot lead to microlensing, since the scale radius $r_s$ is always larger than the lensing tube radius. This conclusion, however, relies on extrapolating the assumed NFW profile and the fitted $c(M)$ relationship to small minicluster masses. At $z=99$ NFW profiles are only resolved for minicluster masses~\cite{Eggemeier:2019khm}:
\begin{equation}
    M>M_{\rm NFW} := 2.59\times 10^{-11}\left(\frac{1\,\mathrm{ meV}}{m_a}\right)^{0.5}M_\odot \, ,
\end{equation}
corresponding to the 30\% of miniclusters with well resolved scale radius at $z=99$ mentioned earlier. Note that in Fig.~\ref{fig:delta_vs_mass}, which shows the distribution at equality, the mass cut between resolved and unresolved scale radius is not so distinct.

Miniclusters with $M<M_{\rm NFW}$ may have steeper PL profiles in their inner regions. This value is redshift dependent and for $z \ll z_{\rm{eq}}$ we expect that $M_{\rm NFW} \sim a$ as in \cref{eq:massGrowth}. The small radius cut-off for a PL profile is given by the axion star radius, which should be smaller than $\xi_c$. 

The above arguments give us three conditions that simulated miniclusters must satisfy in order to be microlensing candidates:
\begin{enumerate}
 \item The effective lensing mass must be greater than the minimum mass for lensing imposed by wave optics effects,
    \begin{equation}
        M_{\rm{eff}} > M_{\rm{min}}\,.
    \end{equation}
    \item The minicluster must allow for the possibility of a steep inner profile, i.e. unresolved NFW scale radius,
    \begin{equation}
        M<M_{\rm NFW}\,.
    \end{equation}
    \item The critical lensing radius must be larger than the estimated axion star radius,
    \begin{equation}
        \xi_c > r_{\star}\,.
    \end{equation}
\end{enumerate}

\begin{figure}[t!]
    \centering
    \includegraphics[width=1\columnwidth]{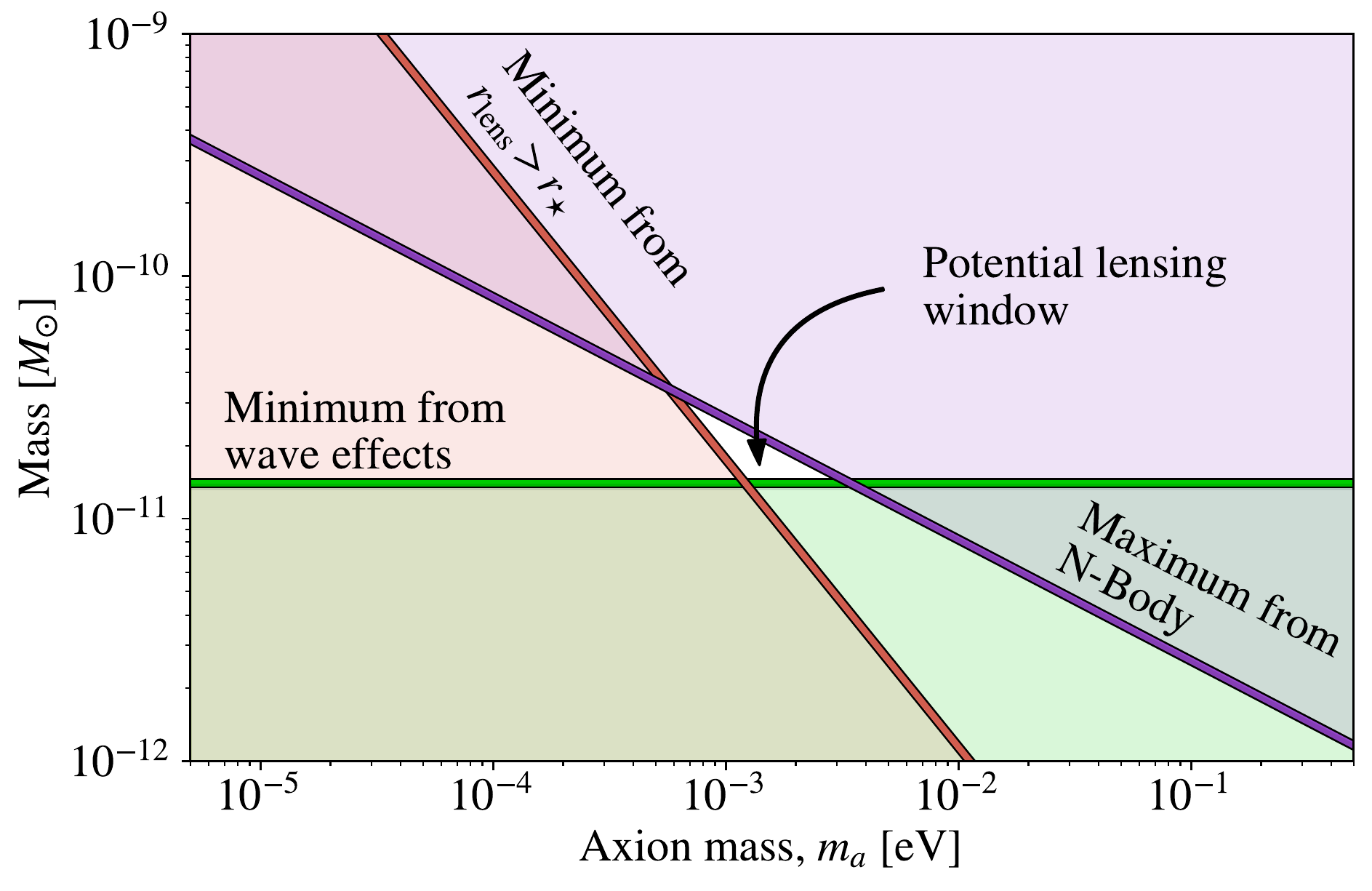}
    \caption{\emph{Conditions for Microlensing}. The potential lensing region, shown by the unshaded region is produced by imposing a minimum halo mass from wave effects (green), a minimum mass from axion stars (red) and a maximum mass from NFW profiles observed in $N$-body simulations (purple).}
    \label{fig:triangle}
\end{figure}

As in Ref.~\cite{Fairbairn:2017sil}, we find the critical lensing radius by solving \cref{eq:extLens,eq:surfPL} and finding the largest radius which gives the same magnification as the outer image for a point mass $\mu_{\rm out} = 1.17$. For a PL minicluster, we find that located halfway between the source and the observer, the minimum impact parameter that produces microlensing above the critical value is given by 
\begin{equation}
    \xi_c = 1.4\times10^{-16}  \left(\frac{M_{200}}{10^{-12}M_{\odot}}\right)^{0.509} \rm{Mpc}\,.
    \label{eq:lenRadius}
\end{equation}
Assuming the core-halo mass relation from Ref.~\cite{Schive2014_PRL} we can set \cref{eq:lenRadius} equal to the radius of the axion star, and demanding $r_\star<\xi_c$ we obtain that microlensing only occurs for miniclusters with masses above
\begin{equation}
    M_{200}>M^{\rm min}_\ast = 1.75\times 10^{-11}\left(\frac{1\,\mathrm{meV}}{m_a}\right)^{1.19}M_\odot \, .
\end{equation}
Calculating the mass bounded within the critical lensing radius, we then find the minicluster's effective lensing mass. For a PL profile, this effective mass is given by
\begin{equation}
    \frac{M_{\rm{eff}}}{M_{200}} = 0.202 \left( \frac{M_{200}}{10^{-12}M_{\odot}} \right)^{0.0175}\,.
    \label{eq:lensMass}
\end{equation}
Taking into account wave optics effects, the effective mass must be greater than $3\times10^{-12}M_{\odot}$ to produce a microlensing signal (see \cref{sec:waveEff}). According to \cref{eq:lensMass}, the virial mass of a PL minicluster must therefore satisfy
\begin{equation}
M_{200}>M_{\rm wave}^{\rm PL}= 1.4\times 10^{-11}\,M_\odot
\label{eqn:wave_optics_PL_halo_mass}
\end{equation}
to be able to lens. 
These conditions are illustrated in Fig.~\ref{fig:triangle}, where we observe that there is a small region of parameter space with $m_a\approx 1\,\mathrm{meV}$ where miniclusters might produce microlensing.

\subsection{Identifying Microlensing Candidates}

Due to the extrapolations required, we will not attempt a statistical microlensing study, but rather we attempt to identify possible microlensing candidates out of the most promising minicluster seeds. We search for lensing candidates from those miniclusters particle tagged using PP as described above and take a uniform sampling in the plane of minicluster mass and initial overdensity as defined at matter-radiation equality. We look to our $N$-body simulations and search for those miniclusters at $z=99$, and compare them to the conditions for microlensing candidates.\footnote{The central mass of the halo is not affected by the assumed halo mass growth ($M\sim a$) from accretion for $z<z_\mathrm{eq}$. Thus, the minicluster capacity for lensing remains unchanged at smaller $z$.} 

\begin{figure}[t!]
\includegraphics[width=1\columnwidth]{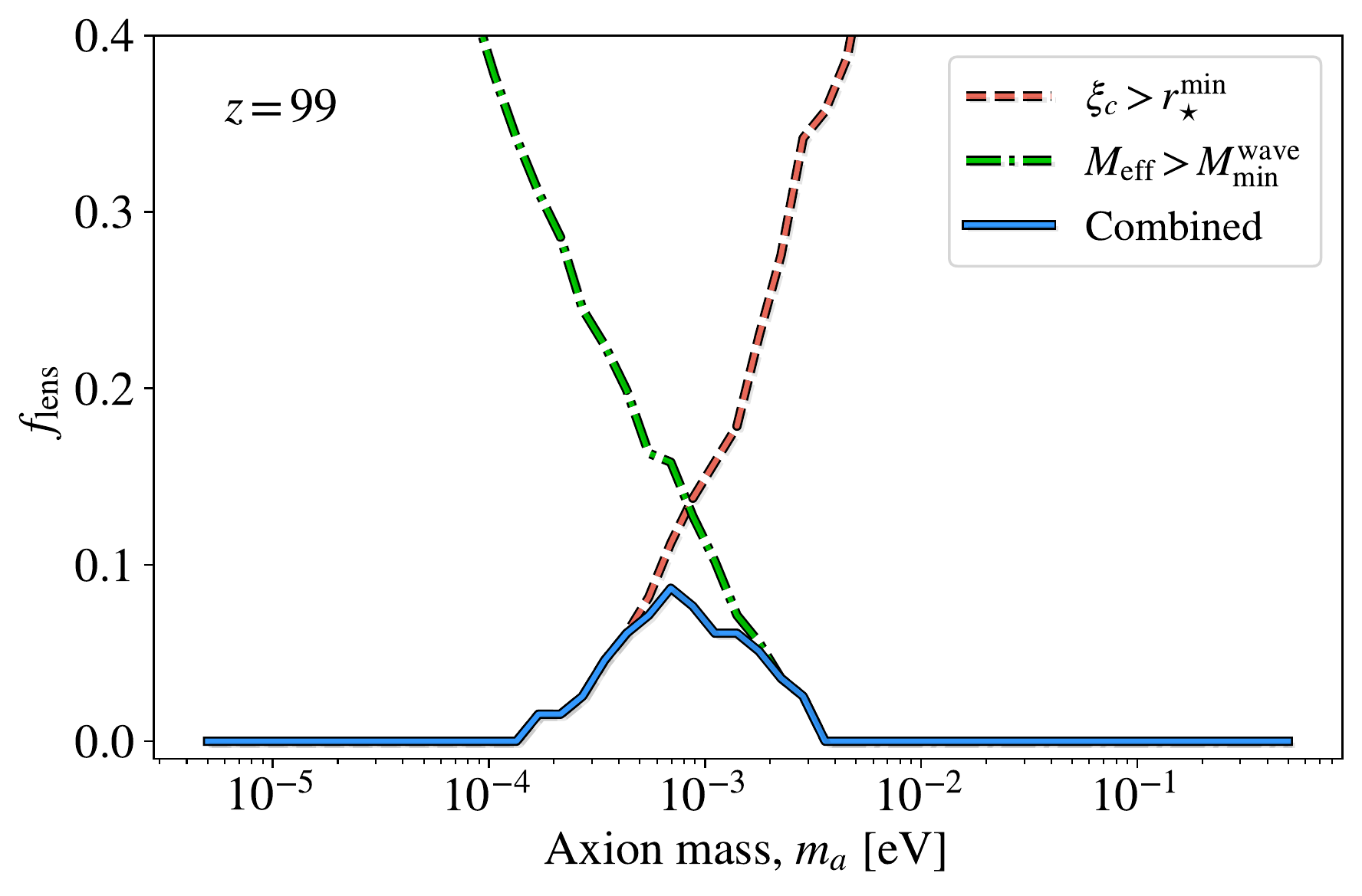}
\caption{\emph{Fraction of samples meeting criteria for lensing}. The fraction was calculated for lenses exactly halfway between the observer and the light source. Lensing and non-lensing halos were shown in Fig.\ref{fig:delta_vs_mass} as yellow stars and orange circles, respectively.
\label{fig:lensFrac}}
\end{figure}

The fraction of objects in the sample found to comply with these conditions is shown as a function of axion mass in Fig.~\ref{fig:lensFrac}. We see that for axion masses $m_a \lesssim 0.2\,\mathrm{meV}$ the predicted radius of the axion star is larger than the critical lensing radius for a halo with a $\rho(r)\sim r^{-2.9}$ power-law profile making microlensing impossible. On the other side, we see that axion masses $m_a \gtrsim 3\,\mathrm{meV}$ produce effective lensing masses that are smaller than the minimum mass required from wave optics effects. However, for axion masses between these two limits, we find that a fraction of our samples might be able to produce microlensing. This fraction peaks at just under 10\% at $m_a \sim 1\,\mathrm{meV}$. We emphasise that this fraction is not representative of the DM mass fraction.

We found the fraction in Fig.~\ref{fig:lensFrac} using the \emph{measured} relationship between $M$ and $v_{\rm vir}$ in order to estimate $R_\star$ (see Fig.~\ref{fig:massesAndradii}), and \emph{not} the core-halo mass relation. Had we applied the core-halo mass relation, then we would have found no microlensing candidates in our sample.

\section{Discussion and Conclusions}\label{sec:discussion}

Axion miniclusters can have important phenomenological implications if they are exceptionally dense. Theoretical arguments suggest that the densest miniclusters form prior to matter-radiation equality, and are characterised by a power-law profile and their initial overdensity. Simulations of miniclusters, on the other hand, resolve a broad mass function from hierarchical structure formation during the matter-dominated era, and large miniclusters with NFW profiles. Do dense minicluster seeds survive this process but remain undetected in simulations e.g. due to selection effects? We have used the Peak-Patch semi-analytical model for halo formation to tag the densest minicluster seeds present at matter-radiation equality in density fields generated from lattice simulations of axion string networks. We then followed the evolution of these seeds in $N$-body simulations. We measured the corresponding minicluster masses, analysed their density profiles, and looked for correlations with their initial overdensity.

A large number of miniclusters have a scale radius too small to be resolved by the available simulations, leaving the door open to the ``dense power-law'' model in part of the mass function. We found no correlation between the minicluster profile and the initial overdensity, although we did find that denser miniclusters were more likely to survive to late times. We furthermore measured the velocity dispersion of the simulated miniclusters, and used this to estimate the radius at which an axion star might form, finding significant scatter in the relation between predicted axion star radius and halo mass.

The conditions for a minicluster to be a microlensing candidate were enumerated  in Section~\ref{sec:lensing_conditions}. The first condition is an absolute limit imposed by the physical observations of the microlensing survey, in this case the HSC Subaru survey. The second and third conditions allow for lensing candidates \emph{within the uncertainties imposed by simulation resolution and methodology} for axion mass $m_a\approx 1\,\mathrm{meV}$. We have only identified the possibility of microlensing candidates, and further simulations are necessary to overcome the resolution issues. Fortunately, our analysis has identified exactly the simulation requirements.

The axion star radius provides one of the relevant cut-offs for microlensing in our analysis, which we estimated by measuring the virial velocity in candidate miniclusters directly in our simulations. If the axion star radius was derived strictly from the core-halo mass relation of Ref.~\cite{Schive2014_PRL} applied to the minicluster mass function, then the axion star radius would be too large to allow for microlensing in  a $10^{-11}M_\odot$ halo for $m_a\approx 1\text{ meV}$. It is the scatter in the relationship between $M$ and $ M_\star\propto v_{\rm vir}$ found in our analysis (see Fig.~\ref{fig:massesAndradii}) that causes us to predict that some miniclusters host smaller axion stars than predicted by the core-halo mass relation, and can thus give rise to observable microlensing amplification. 

In terms of microlensing, the above fact highlights the need for an analysis such as ours to characterise the distribution of miniclusters properties, rather than relying on averages. This also emphasises the observational relevance of possible scatter in the core-halo mass relation~\cite{Chan:2021bja}, with applications elsewhere in axion astrophysics. Ascertaining whether or not scatter in the core-halo mass relation is indeed related to scatter in the $M-v_{\rm vir}$ relation, or other factors, is beyond the scope of the present work.

Our current set of simulations does not resolve axion star formation (since they are $N$-body simulations rather than wave simulations), and only resolves the scale radius in a limited part of the mass function. It is only these unknowns that allow the theoretical wiggle room for the possibility of miniclusters with large enough central densities to microlens. Adaptive zoom-in simulations mixing $N$-body on large scales and wave simulations on small scales~\cite{Veltmaat:2018dfz,Schwabe:2021jne,Eggemeier:2021smj} are required to verify whether such miniclusters can in fact form. The simulations should be run at $m_a\approx 1\,\text{meV}$ and resolve halos with a mass of $10^{-11}M_\odot$ at $z\approx 100$. The radial resolution in these halos should capture any possible NFW scale radius (predicted at $r\approx 1\,\mathrm{AU}$, extrapolating $c(M)$ via the Press-Schechter model) and axion star formation (predicted at $r\approx 10^{-4}\,\mathrm{AU}$).

We have identified microlensing as a probe of miniclusters only for $0.2 \,\mathrm{meV}\lesssim m_a \lesssim 3\,\mathrm{meV}$. If the relic density is dominated by axions produced by string decay, as suggested by recent simulations~\cite{Gorghetto:2020qws}, then $m_a=0.48\,\mathrm{meV}$ to $0.52\,\mathrm{meV}$. Thus, in future, microlensing may be able to confirm or exclude this model for axion production, with important implications for the design of axion direct detection experiments at high-frequency ~\cite{Marsh:2018dlj,Lawson:2019brd,Schutte-Engel:2021bqm,BREAD:2021tpx,Adams:2022pbo}. Miniclusters might also give rise to microlensing in other cosmologies and particle physics models that we have not considered in this work. If there is an early matter-dominated era, as favoured by the solution of the cosmological moduli problem with low-scale SUSY~\cite{Coughlan:1983ci,Acharya:2008bk}, then miniclusters collapse at even earlier times, and are thus denser~\cite{Visinelli:2018wza}. In axion models with domain wall number larger than unity (e.g. Refs.~\cite{DINE1981199,Zhitnitsky:1980tq}), domain wall decay can happen as late as Big Bang Nucleosynthesis~\cite{Hiramatsu:2012sc}, leading to more massive miniclusters requiring much lower average density to lens. Microlensing of miniclusters can thus also act as a probe of these more exotic aspects of axion cosmology.

\section*{Acknowledgements}

DE is supported by the Alexander von Humboldt Foundation, and the German Federal Ministry of Education and Research. DJEM is supported by an Ernest Rutherford Fellowship from the STFC, UK. JR is supported by the grant PGC2018-
095328-B-I00(FEDER/Agencia estatal de investigacion) and
FSE-DGA2017-2019-E12/7R (Gobierno de Aragon/FEDER)
(MINECO/FEDER), the EU through the ITN Elusives H2020-
MSCA-ITN-2015/674896 and the Deutsche Forschungsge-
meinschaft under grant SFB-1258 as a Mercator Fellow. KD acknowledges support by the COMPLEX project from the European Research Council (ERC) under the European Union’s Horizon 2020 research and innovation program grant agreement ERC-2019-AdG 882679 and the Deutsche Forschungsgemeinschaft (DFG, German Research Foundation) under Germany’s Excellence Strategy - EXC-2094 - 390783311. DM thanks Alex Savenkov for thesis work on miniclusters from long-lived domain walls. We thank Igor Tkachev and participants at the QUARKS 2020 conference for useful discussions.

The $N$-body simulations were carried out at the Leibniz Supercomputer Center (LRZ) under the project pr74do. We acknowledge the use of the open source Python packages NumPy \cite{5725236}, SciPy \cite{2019arXiv190710121V} and Matplotlib \cite{4160265}.

\appendix

\section{Numerical Methods}

\subsection{Initial conditions}\label{appendix:ini}

The evolution of the axion field in the post-inflationary scenario comprises:  1) the PQ phase transition, when a global string network forms, 2) a scaling regime during which the string network thins out in a quasi self-similar way, 3) the QCD phase where the QCD potential becomes relevant for the axion zero-modes, builds up domain walls that lead to the fast destruction of the string network, allows axion self-interactions to rearrange the field and finally renders the field non-relativistic at the relevant scales, 4) the free-streaming regime where the axion density field is effective frozen at large scales and small-scale fluctuations stream freely and 5) the gravitational evolution of the axion DM field under gravity forming miniclusters, the object of our study.  

In order to prepare initial conditions for the gravitational collapse phase, we simulated the evolution of a complex scalar field, whose phase is the axion, from the scaling regime, through the QCD phase transition including the free-streaming phase until a redshift of $\sim 10^6$, shortly before gravity starts being relevant for the smallest and densest clumps. 

With current computer limitations, it is not possible to simulate from (1) to (4) and so we start in the scaling regime (2) shortly before the QCD phase (3). We start our simulation with the string density equal to the attractor value found in \cite{Gorghetto:2018myk} a box of comoving length $L=24 L_1$. Here, $L_1$ is the comoving horizon size when the axion zero modes become non-relativistic. This sets the correlation scale of the axion DM density field and thus the typical minicluster mass. Using the temperature dependence of the topological susceptibility $\chi_T(T) = m_a^2(T) f_a^2$ of Ref.~\cite{Borsanyi:2016ksw}, $L_1\equiv (H(t_1)a(t_1))^{-1}=0.0362 (50\,\mu{\rm eV}/m_a)^{0.167}\text{pc}$ was calculated in \cite{Vaquero:2018tib}.
Expressed in terms of $L_1$ and conformal time $\eta_1$, assuming a constant number of degrees of freedom during radiation domination, and fitting $m_a^2\propto (T_1/T)^{-n}=(\eta/\eta_1)^n $, the evolution of the complex-scalar field becomes independent of $f_a$ (and thus of the axion mass), 
\begin{equation}
\Phi''-\Delta \Phi + \frac{m_\Phi^2}{2} \phi(|\Phi|^2-\eta^2) + \eta^{n+3} = 0\,,
\end{equation}
at least to the extent that we can consider $n$ independent of $f_a$. Here, $\Phi$ is the conformal complex scalar, time is normalised to $\eta_1=L_1$, comoving length to $L_1$ and scale factor $a$ to $a_1$. We have used $a/a_1=\eta/\eta_1$. The best fit around values of $m_a\sim 100\mu{\rm eV}$ is $n = 7.6$. Large values behave very similarly so we actually used $n=7$. The independence of $f_a$ means that we can use our results for different values of the axion mass $m_a$. 

The extra parameter $m_\Phi$ is the mass of the radial mode necessary to make axion strings dynamical. We expect $m_\Phi\sim f_a$ but there is a free proportionality constant $\sqrt{2\lambda}$, which could be relatively small. Unfortunately, we cannot simulate with values of $\lambda$ compatible with observations \cite{Gorghetto:2018myk} but one can extrapolate the results to large $m_\Phi$. Current simulations show that the final density field at large scales is quite independent of $m_\Phi$ \cite{Vaquero:2018tib,Buschmann:2019icd} even if the string density grows logarithmically with it \cite{Gorghetto:2018myk}. Therefore we do not perform any extrapolation but simulate with the largest possible values of $m_\Phi$ allowed by our grid $m_\Phi = 1/(\delta L \eta)$ using the Press-Ryden-Spergel trick \cite{Press:1989yh,Moore:1999nt}. We use the public code \url{https://github.com/veintemillas/jaxions} to create the initial conditions, evolve the field through the QCD phase and free-stream until $\eta/\eta_1\sim 10^7$, which corresponds to $z\simeq 10^6$ for $m_a = 100\,\mu\mathrm{eV}$. The simulations were performed on an $8192^3$ lattice.  
We evolve the complex scalar field until the point when the string-network is destroyed by domain walls ($\eta\sim 2.5\eta_1$ in our simulation), then calculate the axion field $\theta={\rm arg}(\Phi)$ and its time derivative and continue the evolution only for the conformal axion field $\psi=\theta a$, neglecting the radial mode, 
\begin{equation}
\label{cthetaevol}
    \psi''-\Delta \psi +\eta^{n+3}\sin(\psi/a) = 0.   
\end{equation}
We stop the simulation when the axitons cannot be possibly resolved anymore ($m_a\delta L a = 1$) at $\eta = 3.7$ and free-stream from there on by solving analytically the linearised version of \cref{cthetaevol} in the WKB approximation until $\eta/\eta_1 = 10^7$. We emphasize that $z_1$ depends on $m_a$ very weakly ($\propto m_a^{0.172}$) and the evolution of the field is only logarithmic at those times. Thus, we can use our results for different axion masses with negligible errors. 

We obtained the axion DM density at $z\sim 10^6$ and sampled it with $1024^3$ particles to study the gravitational evolution in Ref.~\cite{Eggemeier:2019khm} (see also \cref{appendix:nbody} for details).

A few words of caution are in order. Our simulations are far from the desirable values of $m_\Phi$ and produce the observed amount of axion DM for $m_a\sim 20 \mu$eV. We explicitly assume that more physical parameters will change the axion mass required to obtain the correct DM relic abundance, but will not alter significantly the characteristics of the axion density field. 
As $m_\Phi$ increases, larger values of the string density are observed in the scaling regime and, most importantly, a larger spectral index of the axion spectrum radiated by the strings $\propto 1/k^q$. Our simulation stayed below but close to $q\sim 1$. The latest simulations using adaptive-mesh-refinement \cite{Buschmann:2021sdq} found $q=1$ and a detailed study of the scaling regime found that $q$ increases linearly towards $q=1$ without any signs of levelling to $q=1$ \cite{Gorghetto:2020qws}. The value of $q$ at late times determines the total axion yield and the shape of the spectrum. It will determine many of the features of the axion DM density. Our results are expected to be roughly correct if $q$ stays below or close to $1$ even if the total density changes. The axion DM density field might look significantly ($\mathcal{O}(1)$) different if $q\ll 1$.

\subsection{Peak-Patch}\label{appendix:pp}

Our implementation of the Peak-Patch methods broadly follows Ref.~\cite{Stein:2018lrh} and the original paper~\cite{1996ApJS..103....1B}. In contrast to them, we only implement the spherical collapse of a density fluctuation. 
We can split the detailed algorithm into four parts:
\begin{enumerate}
    \item detection of peak candidates
    \item determination of the radius of the collapsing sphere around peak candidates
    \item removal of overlap between halos
    \item calculation of final position and velocity of the final set
\end{enumerate}
The initial density field is generated from simulations similar to the ones presented in Ref.~\cite{Vaquero:2018tib} (see also \cref{appendix:ini} for details). In fact, the same initial conditions as in the $N$-body simulations of Ref.~\cite{Eggemeier:2019khm} are used.

\subsubsection{Detection of peak candidates}
This step has a purely computational sense. One could of course scan the region around each grid cell for collapse at various radii, however, this would be computational overkill. Instead, one can select \enquote{interesting} cells in the first step, called peak candidates. This is done by filtering the density field on various scales and storing all cells that exceed a certain density in the filtered field as peak candidates. We also store the largest filter radius at which a peak candidate was detected. We use a spherical top-hat filter in real-space which translates to a sinc filter in $k$-space. The filter sizes are logarithmically spaced, starting from $2 \Delta x$, where $\Delta x$ is the grid scale. We use about 20 filters, and our threshold density is typically $\delta_c\sim1.5$.

\subsubsection{Determination of the radius of the collapsing sphere around peak candidates}

For each peak candidate, we need to find the maximum radius at which a spherical region around the candidate is collapsed at the redshift in question. In order to do so, we average the density inside a radius $r$ around the peak and check if it is larger than the critical overdensity for collapse. If so, we continue to integrate at an even large $r$. If not, we integrate down to a smaller $r$. If we find a transition between collapse and uncollapsed between two radial bins, we interpolate the radius linearly between the bins. The mass can be obtained from this as the mass of a sphere with said radius at average cosmic density. In contrast to Ref.~\cite{1996ApJS..103....1B}, we use a radial binning that allows fractional contributions from cells that are partially inside the integration radius, which increases the precision at the lowest radii/masses slightly.

\subsubsection{Removal of overlap between halos}
The first step in the overlap removal is to remove all halos whose centre falls within the radius of a larger halo. This has to be done hierarchically, that is, we have to check for overlapping halos in the most massive halo first and proceed into the less massive (surviving) hosts. To see why, consider three halos with decreasing mass, A, B, and C, with A containing Bs centre, B containing Cs centre, A not containing Cs centre. Depending on the order of operation, C survives or not. 

The second step is to approximately resize halos that still overlap to account for the mass that counted twice in the overlap region. We use half-exclusion here, as described in Ref.~\cite{Stein:2018lrh}, and reduce both the mass of the lower and higher mass halo.

\subsubsection{Calculation of final position and velocity of the final set}
For the last step, one needs to calculate the displacement field from the density field. The equations are given in Ref.~\cite{Stein:2018lrh}. We support both the linear approximation and the second-order correction. Note that the memory consumption doubles when going from first to second-order, as the second-order displacement contribution adds 3 new floats per lattice site. At the present, we do not require the halo positions from PP to be accurate, and so we use only first-order displacements.

\subsection{$N$-body simulations}
\label{appendix:nbody}
The initial axion density distribution of the $N$-body simulations from Ref.~\cite{Eggemeier:2019khm} was generated by large lattice simulations~\cite{Vaquero:2018tib} evolving the axion field from PQ symmetry breaking until it becomes non-relativistic (see also \cref{appendix:ini} for details). A conversion of this axion distribution at redshift $z\simeq 10^6$ into $1024^3$ particles with a mass of $2.454\times 10^{-17}\,M_\odot$ and initial velocities set to zero serves as the initial conditions for the succeeding $N$-body simulations. 

They were performed with the OpenMP/MPI optimized \textsc{Gadget-3} code which is a predecessor of the recently published \textsc{Gadget-4}~\cite{Gadget4}. The limit of the spatial resolution is determined by the numerical softening length which was set to $1\,\mathrm{AU}/h$ in comoving units. Considering only gravitational interactions among the particles, they were evolved with a comoving box side length of $L=0.864\,\mathrm{pc} = 24 L_1$ (see Appendix \ref{appendix:ini}) for an axion mass of $m = 50\,\mu\mathrm{eV}$ until a final redshift of $z=99$. At this time, the scales that correspond to the length of the simulation box become nonlinear, so simulations for redshifts smaller than $z=99$ can only be trusted when larger box sizes are considered. 

Standard $\Lambda$CDM parameters $\Omega_{m,0} = 0.3$, $\Omega_{r,0} = 8.486\times 10^{-5}$ (taking into account photons and three massless neutrino species), $\Omega_{\Lambda,0} = 0.7$ and $H_0 = 100h\,\mathrm{km}\,\mathrm{s}^{-1}\,\mathrm{Mpc}^{-1}$ with $h=0.7$ were used to evolve the Hubble parameter, 
\begin{align}
    H(z) = H_0\left(\Omega_{m,0}(1+z)^3 + \Omega_{r,0}(1+z)^4 + \Omega_{\Lambda,0}\right)^{1/2}\,.
\end{align}

Defining an axion minicluster as a collection of gravitationally bound particles, the \textsc{Subfind} halo finder~\cite{Subfind,Dolag2009} was used to identify halos and their sub-halos with a minimum number of 32 and 20 bound particles, respectively. The centre of a halo is determined by the position of the minimum of its gravitational potential. Their size is set by the virial radius $r_\mathrm{vir}$, at which the enclosed average density of a halo equals the virial parameter $\Delta_\mathrm{vir}$ times the critical density $\rho_c = 3H^2/(8\pi G)$. The virial mass of a minicluster is then $M_\mathrm{vir}=4\pi/3 \Delta_\mathrm{vir}\rho_c r_\mathrm{vir}^3$. The virial parameter is given by 
\begin{align}
    \Delta_\mathrm{vir}/\Omega_m(z) = 18\pi^2 + 82(\Omega_m(z)-1) - 39(\Omega_m(z)-1)^2\,,
\end{align}
where 
\begin{align}
    \Omega_m(z) = \frac{\Omega_{m,0}(1+z)^3}{\Omega_{m,0}(1+z)^3 + \Omega_{r,0}(1+z)^4 + \Omega_{\Lambda,0}}\,.
\end{align}
In this convention, the redshift-dependent value of the virial parameter depends on the cosmology. Instead, it is also quite common to take $\Delta_\mathrm{vir}=200$ to define virial quantities of halos.

\section{Miniclusters from Self-Similar Infall}
\label{sec:infall}

We summarise here the theory behind power-law miniclusters from spherical collapse and self-similar infall. 

While larger miniclusters gravitationally collapse at later times and exhibit smaller initial axion density fluctuations, miniclusters originating from a large initial overdensity already collapse deep in the radiation-dominated epoch. 
As was shown in Ref.~\cite{Eggemeier:2019khm}, the gravitational evolution of axion miniclusters after matter-radiation equality is dominated by mergers of smaller miniclusters into larger ones. 

It is expected that less concentrated objects are more likely to be tidally disrupted by these merger events. However, an early distribution of miniclusters with sufficiently large enough initial overdensity can end up as substructures within larger miniclusters without being significantly affected by tidal encounters. 
Thus, their initial density profiles and concentrations can be assumed to remain unchanged.
They form primarily via self-similar infall, and analytical studies typically predict that this should result in power-law density profiles of the form~\cite{1975ApJ...201..296G, 1985ApJS...58...39B}
\begin{equation}
    \rho(r) = \rho_0\left( \frac{r}{r_0} \right)^{-\alpha}\,.
    \label{eq:pl_dens}
\end{equation}
It was first realised that the self-similar infall of an initially static uniform spherical overdensity produces a density profile with $\alpha = 3$. However, later studies showed that when matter which is initially expanding instead produces a density profile $\alpha = 9/4$~\cite{1975ApJ...201..296G, 1985ApJS...58...39B}. 

An axion field with a white-noise initial power spectrum simulated with $N$-body until $z=3000$, produced halos with this predicted density profile \cite{Zurek:2006sy}. This has therefore been commonly assumed to be the density profile for axion miniclusters \cite{Fairbairn:2017dmf, Fairbairn:2017sil, OHare:2017yze}. 

To fully predict the density profile, we also need a relationship between the mass and radius of these miniclusters. One approach used in the past has been to take the total average density of the minicluster to be~\cite{Kolb:MCdens,Fairbairn:2017sil, Kavanagh:2020gcy}
\begin{equation}
    \langle \rho \rangle =  140\rho_{\mathrm{eq}}\delta_i^3(1+\delta_i)\,,
\end{equation}
where $\delta_i$ is the initial overdensity and $\rho_{\mathrm{eq}}$ is the density of the universe at matter-radiation equality. This formula is found by considering the spherical collapse of a tophat overdensity including the internal mass contribution from radiation \cite{Kolb:MCdens}. We can use this to calculate the radius for a halo of some mass and initial overdensity $\delta_i$.

If we define $r_0$ to be this radius, we can integrate \cref{eq:pl_dens} with respect to the radius to find that 
\begin{equation}
    M = \frac{4\pi\rho_0}{3-\alpha}r_0^3,
    \label{eq:PLradius}
\end{equation}
and hence
\begin{equation}
    \rho_0 = \frac{3-\alpha}{3}\langle \rho \rangle\,.
\end{equation}
Now, given a mass and an initial overdensity $\delta_i$, we can predict a power-law profile. However, at each point in time, there is a single threshold overdensity that defines which objects have collapsed. Therefore, all virialised objects at a single point in time can be considered to have the same initial overdensity. By comparing the linear growth of an adiabatic perturbation overdensity with the non-linear collapse of a spherical tophat, we find that the initial overdensity which collapses at redshift $z$ is given by
\begin{equation}
    \delta_{i, \mathrm{col}}(z) = \frac{1.686}{1+D(z)}\,,
    \label{eq:init_del}
\end{equation}
where $D(z)$ is the growth factor
\begin{equation}
    D(z) = 1 + \frac{3}{2} \left(\frac{z_{\mathrm{eq}} + 1}{z + 1} \right)\,,
\end{equation}
and $z_{\mathrm{eq}}$ is the redshift at matter-radiation equality. The collapsing initial overdensity is therefore inversely proportional to the scale factor and $\delta_{i, \mathrm{col}}(z_{\mathrm{eq}})~\sim 1$ \cite{Ellis:2020gtq}.

\section{Axion Stars}\label{appendix:star}

The density profile of a minicluster has a natural \enquote{UV cut-off} on small scales caused by the presence of a central axion star. This scale was used in our microlensing analysis and model of the minicluster density profile. Axion stars are not resolved in our $N$-body simulations. However, we can use known properties of halos with axion stars to ``paint on'' the stars to our profiles. We describe these properties here.

Since the axion likely has a very small mass, it will also have a comparatively large de Broglie wavelength. This suppresses structure formation on small scales due to the uncertainty principle. Additionally, when the axions are cold, they can form a Bose-Einstein condensate. This condensation produces a solitonic core known as an axion star. This gives a flat core at the centre of dark matter halos~\cite{Hu:2000ke,Schive2014_Nature,Eggemeier:2019jsu,Niemeyer:2019aqm}. 

These solitons are most often considered for ultra-light axions ($m_a \sim 10^{-22}$eV) as a way of explaining problems faced by classic CDM such as the missing-satellite \cite{Klypin:1999uc, Moore:1999nt} and cusp-core problems \cite{Dubinski:1991bm, Navarro:1995iw}. However, it has been shown that they also form for masses which are relevant for the QCD axion \cite{Eggemeier:2019jsu}.

Axion stars make up a significant part of the structure of miniclusters. This is particularly valid for the smaller halos and sub-halos since, somewhat counter-intuitively, the radius of an axion star is inversely proportional to the mass of its host halo.

Using wave simulations, Schive et al. uncovered the following relation between axion star core mass $M_{\rm{*}}$ and their host halos $M_{\rm{h}}$~\cite{Schive2014_PRL},
\begin{equation}
    M_{*} \propto a^{-1/2} M_{\rm{h}}^{1/3}.
\end{equation}
The axion star mass can be conveniently expressed in terms of the virial velocity of its parent halo. If the axion star is in virial equilibrium with the halo, simulations have found this relation to be
\begin{equation}
    M_\star = 4.69 \frac{\hbar}{m_a}\frac{v_{\mathrm{vir}}}{G}\,,
    \label{eq:axionstar_mass_velocity}
\end{equation}
where $v_{\mathrm{vir}}$ is the virial velocity of the parent halo~\cite{Eggemeier:2021smj}. 
The radial density profile of an axion star is well approximated by~\cite{Schive2014_Nature,Schive2014_PRL}
\begin{equation}
    \rho_\star(r) \simeq \rho_0 \left[1 + 0.091\left( \frac{r}{r_\star} \right)^2\right]^{-8}\,, \label{eqn:soliton_profile}
\end{equation}
where $r_\star$ denotes the axion star half mass radius and $\rho_0$ its central (physical) density 
\begin{equation}
    \rho_0 \simeq 7.8\times 10^{-14} \left(\frac{50\,\mu\mathrm{eV}}{m_a}\right)^2\left( \frac{10^{-3} \text{pc}}{r_\star}\right)^4 \frac{M_{\odot}}{\text{pc}^3}\,.
\end{equation}

\section{Gravitational Microlensing}
\label{appendix:micro}

\subsection{Basic microlensing}
\begin{figure}[t]
\includegraphics[width=1\columnwidth, trim=0 0 0 0, clip]{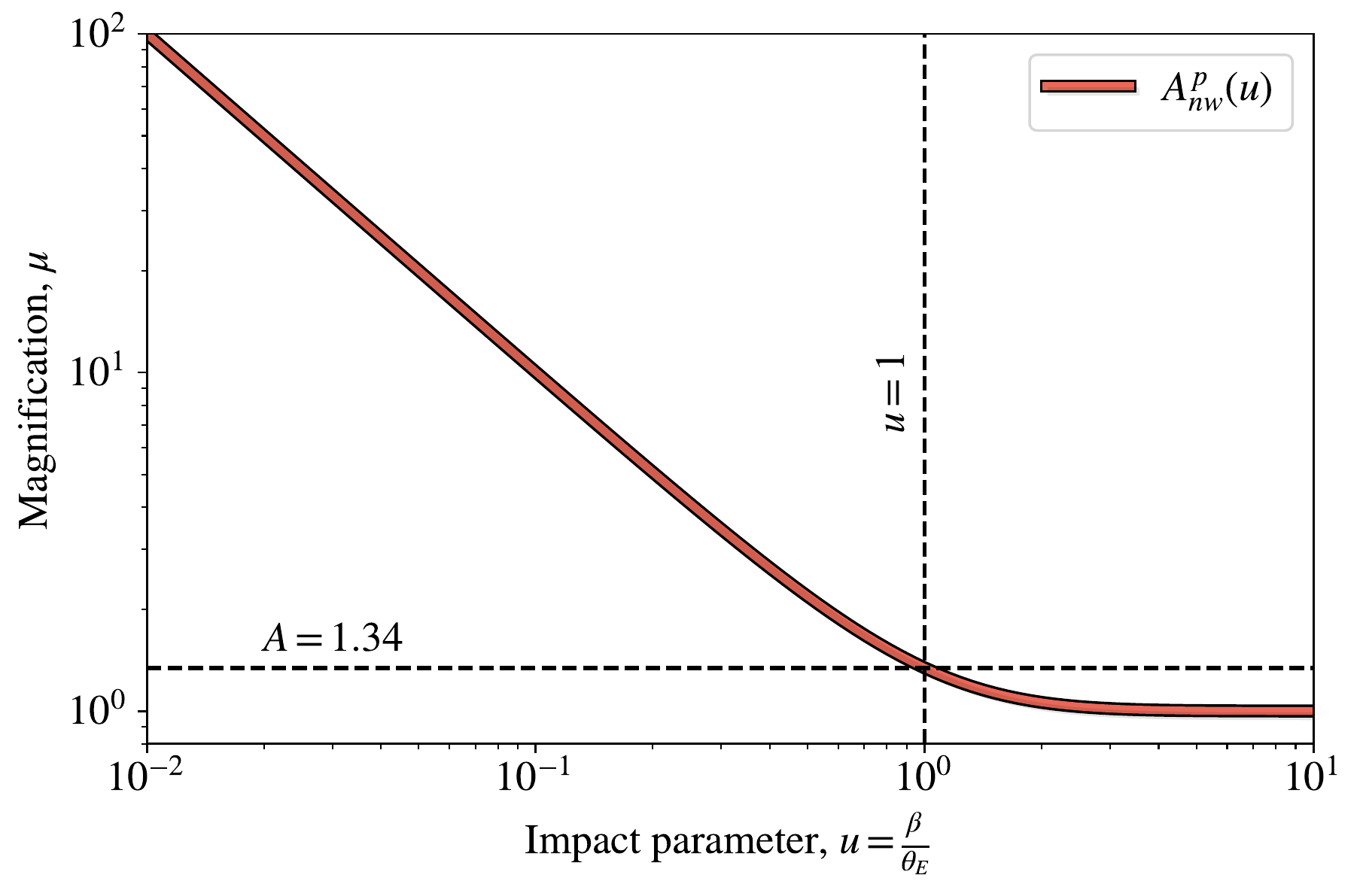}
\caption{\emph{Point mass magnification.} Magnification due to lensing by a point mass as a function of impact parameter under the geometrical optics approximation.} 
\label{fig:noWaveMag}
\end{figure}

The magnification for a point source by a point mass, such as a primordial black hole (PBH), under the geometrical optics approximation is
\begin{equation}
    \mu^p_{nw}(u) = \frac{u^2 + 2}{u\sqrt{u^2 + 4}}\,,
    \label{eq:noWave}
\end{equation}
where $u = \beta/\theta_E$ is a dimensionless impact parameter. We see, as shown in \cref{fig:noWaveMag}, that the magnification is greater than $1.34$ for $u<1$. From this, we define a threshold impact parameter $u_{\textsc{t}}$ as the maximum impact parameter which produces a magnification of 1.34. We will assume that only objects which cross the lensing tube with an impact parameter less than this threshold can produce measurable lensing events. For point masses under this approximation, the threshold impact parameter is one by definition. However, as we will see later, for more complex lensing models this impact parameter is dependent on both the lens structure and its distance from the observer.

The differential detection rate is given by
\begin{equation}
    \frac{\D \Gamma_{\mathrm{PBH}}}{\D \hat{t}} = 2\frac{\Omega_\mathrm{PBH}}{\Omega_\mathrm{DM}} D_S \int^{1}_0 \D x \int ^ {u_T} _ 0 \D u_\mathrm{min} X\,,
\end{equation}
where
\begin{equation}
    X = \frac{v^4}{\sqrt{u_T^2 - u_\mathrm{min}^2}} \frac{\rho_{\mathrm{DM}}(x)}{M_\mathrm{PBH} v_c^2(D_L)}  \exp{\left[ -\frac{v^2}{v_c^2(x)} \right]}\,
\end{equation}
and 
\begin{equation}
    v = 2 \frac{ R_E\sqrt{u_T - u_\mathrm{min}^2} } { \hat{t}}\,,
\end{equation}
where $D_L$ is the distance to the lens, $D_S$ is the distance to the source, $x = D_L/D_S<1$, and $\hat t$ denotes the duration of the lensing event.
The number of observed events expected is given by
\begin{equation}
    N_{\mathrm{exp}} = E \int ^{\infty}_0 \frac{\mathrm{d}\Gamma}{\mathrm{d}\hat{t}} \epsilon(\hat{t})\mathrm{d}\hat{t}\,,
\end{equation}
where $E$ is the exposure and $\epsilon(\hat{t})$ is the probability of the lensing event being observed, i.e. the ``lensing efficiency''.

\subsection{Extended lenses}
The magnification of an extended lens is given by
\begin{equation}
    \mu(\xi) = \frac{1}{(1-B)(1 + B - C)}\,,
    \label{eq:extLens}
\end{equation}
with parameters
\begin{equation}
    C(\xi) = \frac{1}{ \Sigma_c \pi \xi } \frac{\D M(\xi)}{\D\xi}, \;
    B(\xi) = \frac{M(\xi)}{ \Sigma_c \pi \xi^2 }, \;
    \Sigma_c = \frac{c^2D_S}{4\pi G D_L D_{LS}},
    \label{eq:extLens2}
\end{equation}
with $D_{LS}=D_S-D_L$ the distance from the lens to the source. We can then calculate the maximum lensing radius $\xi_\mathrm{max}$ defined to be the maximum radius to give $\mu = 1.17$. From this, we can calculate the lensing ratio
\begin{equation}
    \mathcal{R} = \frac{\xi_\mathrm{max}}{R_E}\,,
\end{equation}
which we average over the lens positions from observer to source.

\subsubsection{NFW profiles}

Integrating the NFW profile as given by \cref{eq:nfw_profile} along the line of sight from $-\infty$ to $\infty$, it can be shown that the NFW profile has a surface mass density given by
\begin{equation}
    \Sigma(x) = \frac{\rho_sr_s}{x^2 -1}f(x)\,,
    \label{eq:SigNFW}
\end{equation}
in which
\begin{equation}
    f(x) = 
    \begin{cases} 1 - \frac{2}{\sqrt{x^2 - 1}} \arctan{\sqrt{ \frac{x-1}{x+1} }} &\mbox{if } x> 1\,,\\
    1 - \frac{2}{\sqrt{1 - x^2}} \arctan{\sqrt{ \frac{1-x}{1+x} }} &\mbox{if } x < 1\,,\\
    0 &\mbox{if } x = 1\,.
    \end{cases}
\end{equation}
Here, $x = \xi/r_s$ and $\xi$ is the radius in the lens plane~\cite{MeneghettiBook}. The surface mass profile can then be calculated numerically by integrating \cref{eq:SigNFW} as
\begin{equation}
    M(\xi) = 2 \pi \int_0^{\xi} \Sigma\left(\frac{\xi}{r_s}\right) \xi \,\D\xi\,.
    \label{eq:NFWmassProf}
\end{equation}
It should be noted that we have assumed that the halo extends infinitely beyond its virial radius which we know to be untrue. However, the mass contributed by large radii is very small. Additionally, we can make a partial fix to this approximation by enforcing that  $M(r>R_{\rm{vir}})=M_{\rm{vir}}$.

\begin{figure}
\includegraphics[width=\columnwidth, trim=0 20 0 21, clip]{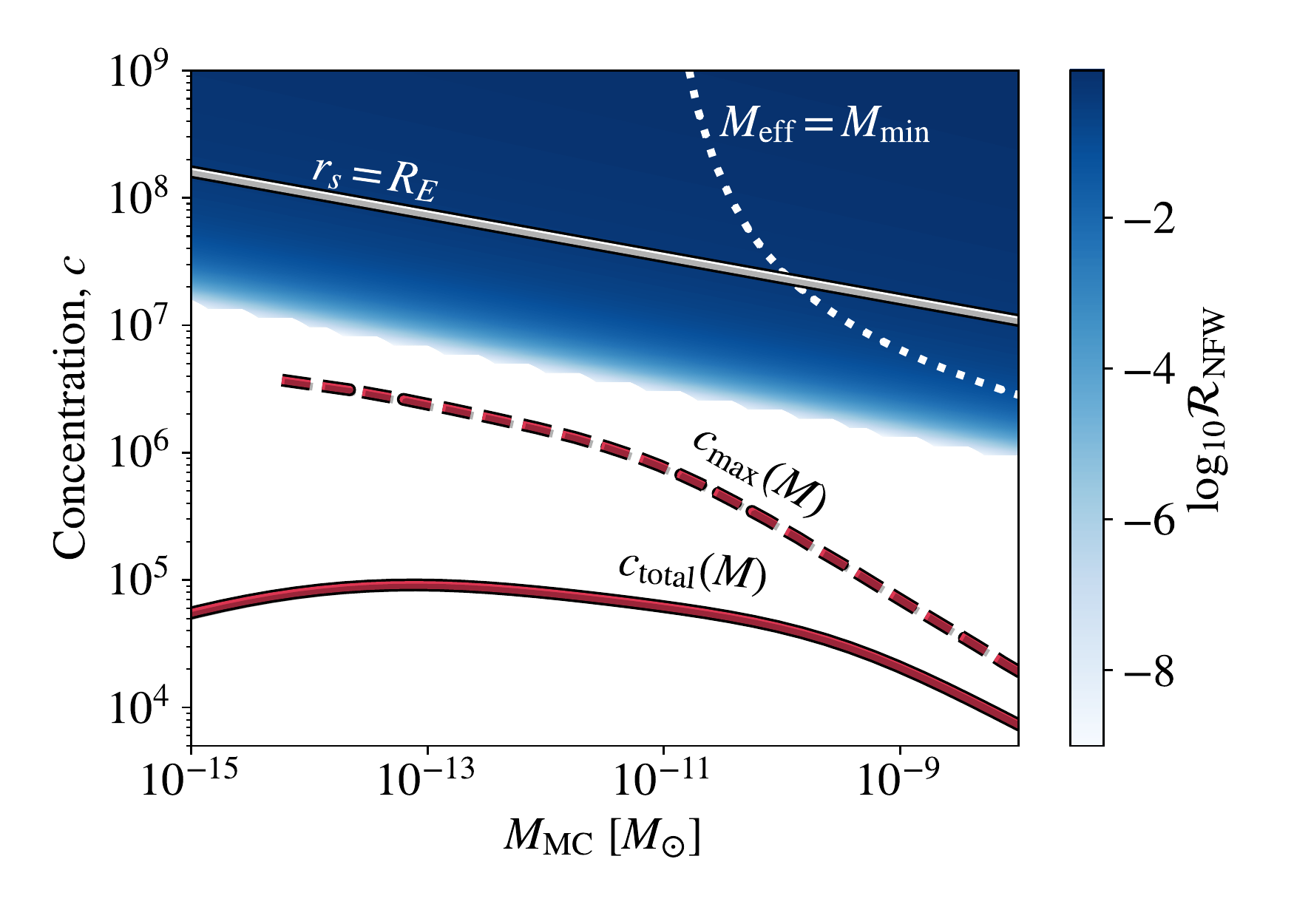}
\caption{\emph{Relative lensing tube size $\mathcal{R}$ for NFW profiles}. White regions indicate where $\mathcal{R}=0$ identically, due to the shallow inner slope $r^{-1}$ preventing caustics. The effective mass within the tube, $M_{\rm eff}$, should exceed $M_{\rm min}$ to avoid wave optics effects suppressing magnification. NFW miniclusters with $c(M)$ calibrated to $N$-body simulations at $z=99$ and extrapolated to $z=0$ have $\mathcal{R}=0$ and $M_{\rm eff}\ll M_{\rm min}$ and thus cannot lens. This is the case for the ``maximum'' concentration, $c_{\rm max}$, discussed in the text. }
\label{fig:NFWlens}
\end{figure}

We can then substitute \cref{eq:NFWmassProf} into \cref{eq:extLens} to calculate the magnification as a function of impact parameter. We can do this as a function of concentration $c$ and minicluster mass $M_{\textsc{mc}}$ as shown in Fig.~\ref{fig:NFWlens}. 

We computed $\mathcal{R}$ and the effective lens mass $M_{\rm eff}$ for NFW profiles. In order for NFW profiles to lens, they must have exceptionally high values of $c\gtrsim 10^7$ such that $M_{\rm eff}$ exceeds the minimum for wave optics, and $\mathcal{R}>0$. Minicluster halos described by the $c(M)$ relationship found above do not lead to any microlensing signal, since $\mathcal{R}=0$ along the extrapolated $z=0$ concentration curve as shown in Fig.~\ref{fig:NFWlens}. 

We can also estimate a maximum possible concentration $c_{\mathrm{max}}(M)$ that halos, or subhalos, could have in the NFW model assuming Press-Schechter. This is done by noting the earliest redshift at which any halos of each mass can form. Therefore, we can estimate the maximum concentration any surviving halo could have by taking this earliest formation redshift to be the collapse redshift in \cref{eq:NFWConc1}. Even using $c_{\mathrm{max}}(M)$, we find that NFW halos are unable to lens.
These results are illustrated in Fig.~\ref{fig:NFWlens} for the axion mass $m_a=50\,\mu\text{eV}$. 
It is possible to rescale these results to a lower axion mass, thus moving the $c(M)$ curve to the right and closer to the $\mathcal{R}>0$ region. However, the $M_{\rm eff}$ region would be moved by exactly the same amount. Therefore, we see that it is impossible to achieve microlensing on an NFW minicluster by considering a different axion mass.

%%%%%%%%%%%%%%%%%%%%
%%%%%%%%%%%%%%%%%%%

\subsubsection{Power-law profiles}
Substituting the power-law profile given by \cref{eq:pl_dens} into the Abel integral, one obtains the surface mass density 
\begin{equation}
    \Sigma_\mathrm{PL}(\xi) = \rho_0 r_0^{-\alpha} B \left( \frac{\alpha}{2} - \frac{1}{2}, \frac{1}{2} \right) \xi^{1-\alpha}\,.
\end{equation}
Here, $B(x,y)$ is the Beta-function which is defined in terms of the Gamma-function $\Gamma(x)$ by
\begin{equation}
    B(x, y) = B(y, x) = \frac{ \Gamma(x) \Gamma(y)  }{ \Gamma(x + y) }\,.
\end{equation}
Integrating the radial density from zero to the virial radius, we find that
\begin{equation}
    \rho_0 r_0^{\alpha} = \frac{3-\alpha}{4\pi} \frac{M_\mathrm{vir}}{R_ \mathrm{vir}^{3-\alpha}}\,,
\end{equation}
allowing us to define the surface mass density in terms of the viral mass and radius as 
\begin{equation}
    \Sigma_\mathrm{PL}(\xi) = \frac{3-\alpha}{4\pi} \frac{M_\mathrm{vir}}{R_\mathrm{vir}^{3-\alpha}} B \left( \frac{\alpha}{2} - \frac{1}{2}, \frac{1}{2} \right) \xi^{1-\alpha}\,.
    \label{eq:surfPL}
\end{equation}
We see that this can be straightforwardly integrated to give the surface mass profile

\begin{equation}
    M_\mathrm{PL}(\xi) = \frac{1}{2} \frac{M_\mathrm{vir}}{R_\mathrm{vir}^{3-\alpha}} B \left( \frac{\alpha}{2} - \frac{1}{2}, \frac{1}{2} \right) \xi^{3-\alpha}\,.
\end{equation}

\begin{figure}
\includegraphics[width=\columnwidth, trim=0 20 0 0, clip]{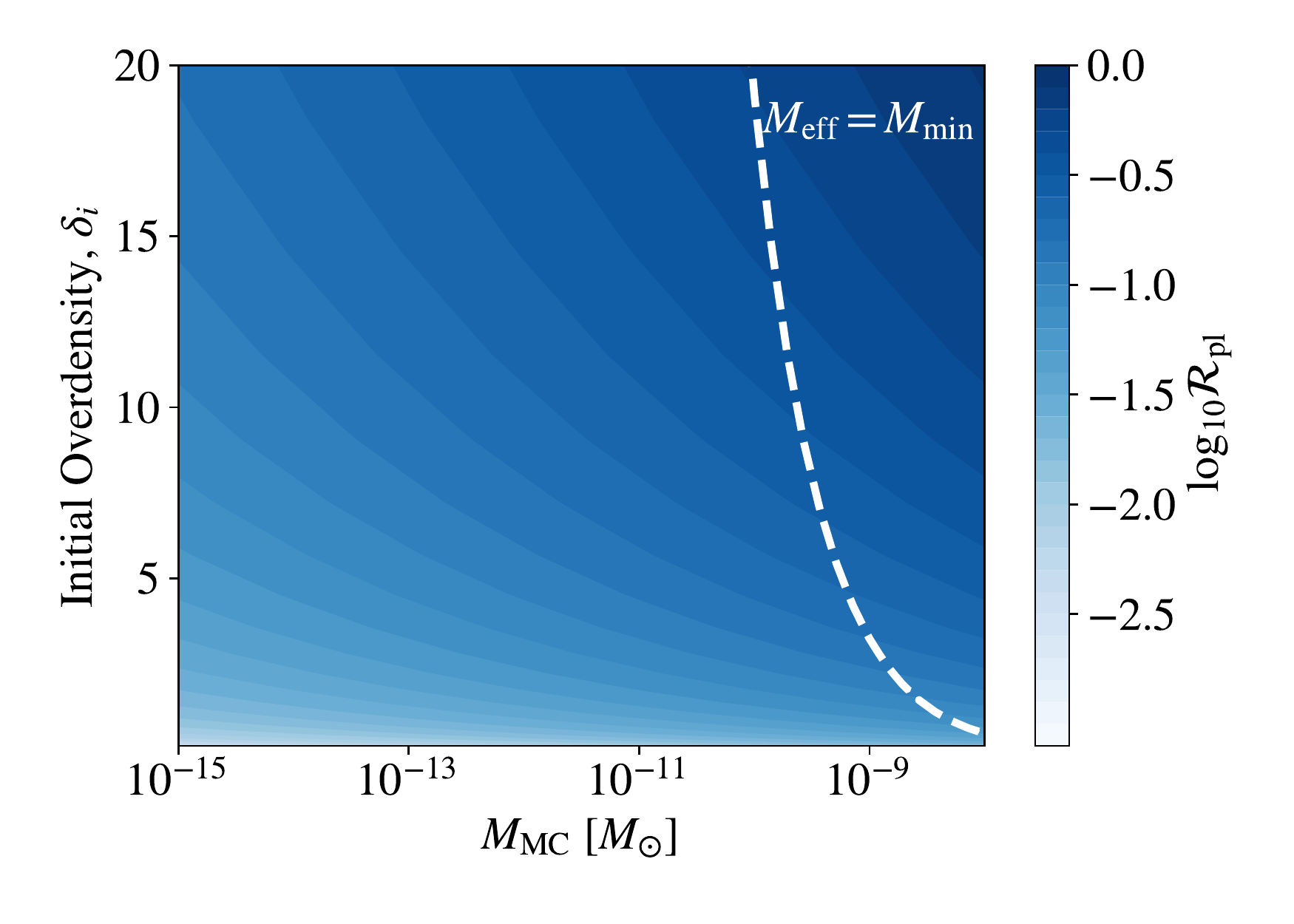}
\caption{\emph{Relative lensing tube size $\mathcal{R}$ for power-law profiles}. The colour map shows the lensing tube radius in units of the Einstein radius for a power-law $\rho\propto r^{-2.9}$. The dashed line sets the enclosed mass within the lensing tube equal to be larger than the minimum mass accounting for wave optics effects. In this figure, the average overdensity is fixed from the initial overdensity, $\delta_i$, assuming spherical collapse.}
\label{fig:PLlense}
\end{figure}

\subsection{Wave optics effects}
\label{sec:waveEff}
The ``geometrical optics approximation'' outlined above is only applicable when the Schwarzschild radius of the gravitational lenses is much larger than the wavelength of the lensed light. For point-like lenses, this simply means that the approximation is only valid for large masses. For smaller masses, this approximation breaks down and we have to consider wave optics effects. As we will see, these wave effects suppress the magnification of light due to small masses, therefore, reducing their detectability. 

When wave optics are included, the magnification due to a point mass is given by
\begin{equation}
	A^p_w(w,u) = \frac{\pi w}{1 - e^{-\pi w}} \left | _1F_1 \left( \frac{i}{2}w, 1;\frac{i}{2}wu^2  \right)   \right|^2\,,
\end{equation}
where again we follow the notation used by Ref.~\cite{Sugiyama:2019dgt} in which ``$w$'' denoted ``wave-effect'' and $_1F_1$ is the confluent hypergeometric function. In the limit of small wavelengths ($w\gg1$) we recover the magnification under the geometrical optics approximation
\begin{equation}
	\mu_{geo}^p(w, u) = \frac{u^2 + 2}{u\sqrt{u^2 + 4}} + \frac{2}{u\sqrt{u^2 + 4}} \sin U(u)\,,
\end{equation}
where
\begin{equation}
    U(u) = w \left( \frac{1}{2}u\sqrt{u^2 + 4} +  \log\left | \frac{\sqrt{u^2 + 4} + u}{\sqrt{u^2 + 4} - u} \right| \right)\,.
    \label{eq:waveMag}
\end{equation}
We see that we recover the ``no-wave'' magnification of \cref{eq:noWave} plus an additional contribution that rapidly oscillates with $u$. Using Eq.~\cref{eq:waveMag}, we find that the central magnification $\mu_{geo}^p(u = 0)$ decreases with lens mass. This means that below some threshold lens mass $M_{\rm min}$, the maximum magnification will be less than the threshold value of 1.34. Hence, lenses smaller than $M_{\rm min} \approx 3 \times 10^{-12} M_{\odot}$ are unable to produce a microlensing signal for a typical r-band filter wavelength of $\lambda \approx 6000\, \mathrm{\AA}$.

\section{Scaling relations}

In this work we used the results of Refs.~\cite{Vaquero:2018tib,Eggemeier:2019khm} which were computed at $m_a= 50\,\mu\text{eV}$. For our Peak-Patch and semi-analytic models, we rescaled the initial data of Ref.~\cite{Vaquero:2018tib} to different axion masses using the following scaling relations. 
The masses of miniclusters scale as
\be
    M \propto \left( \frac{1}{m_a}\right)^{0.5}\,.
\ee
As a result of this, length scales behave similarly as
\be
    L \propto \left( \frac{1}{m_a}\right)^{0.167}\,.
\ee
We can calculate similar scalings for our predicted axion star masses and radii. We expect that the virial velocity should scale as 
\be
    v_{\rm{vir}} \propto M^{1/3} \sim \left(\frac{1}{m_a}\right)^{1/6}\,.
\ee
Therefore, from \cref{eq:axionstar_mass_velocity}, we see that the axion star mass will scale as
\be
    M_{\star}\propto \frac{v_{\rm{vir}}}{m_a} \sim \left(\frac{1}{m_a}\right)^{7/6}\,.
\ee
Finally, we can also use this to find that the axion star radius behaves as
\begin{equation}
    r_{\star}\propto \frac{1}{M_{\star}m_a^2} \sim
    \left(\frac{1}{m_a}\right)^{5/6}\,.
\end{equation}
We see that as the axion mass increases, the halo masses and radii decrease. The axion stars, however, have additional dependencies on the axion mass. We see that increasing axion mass increases the axion star mass while decreasing the axion star radius. Importantly, the axion star radius shrinks quicker than the parent halo. Therefore, at larger axion masses, the relative size of the axion star to the halo is smaller. 

\bibliographystyle{h-physrev3}
\bibliography{axion}

\end{document}